\documentclass[a4paper,12pt]{article}
\usepackage{amsfonts, amsmath, amssymb, amsthm}
\usepackage[blocks]{authblk}
\usepackage[english]{babel}
\usepackage{booktabs}
\usepackage{bm}
\usepackage{bbm}
\usepackage{cases}
\usepackage{eucal}
\usepackage{enumerate}
\usepackage{enumitem}
\usepackage{float}
\usepackage{dsfont}
\usepackage{epsfig}
\usepackage{fontenc}
\usepackage[hmargin = 1.6cm, vmargin = 3cm]{geometry}
\usepackage[colorlinks, allcolors = blue]{hyperref}
\usepackage{lscape}
\usepackage{mathrsfs}
\usepackage{mdframed}
\usepackage{natbib}
\usepackage{rotating}
\usepackage{subfig}
\usepackage{verbatim}
\usepackage[table]{xcolor}

\usepackage{color}

\newtheorem{theorem}{Theorem}

\newtheorem{example}{Example}

\usepackage{url}
\usepackage{amsmath,amsfonts,amssymb,amscd,mathtools,bm,bbm}
\usepackage{multirow,tabularx,subfig,graphicx}
\usepackage{tikz}
\usetikzlibrary{calc,trees,positioning,arrows,chains,shapes.geometric, decorations.pathreplacing,decorations.pathmorphing,shapes, matrix,shapes.symbols}
\tikzset{ >=stealth', punktchain/.style={ rectangle, rounded corners, draw=black, very thick, text width=10em, minimum height=3em, text centered, on chain}}

\newcommand{\norm}[1]{\left\lVert#1\right\rVert}

\newcommand{\gammaV}{\mbox{\boldmath \(\gamma\)}}

 \let\oldthebibliography=\thebibliography
 \let\oldendthebibliography=\endthebibliography
 \renewenvironment{thebibliography}[1]{%
   \footnotesize
   \oldthebibliography{#1}%
   \setlength{\parskip}{0mm}%
   \setlength{\itemsep}{0mm}%
 }{\oldendthebibliography}

\newcommand{\dif}{\ensuremath{\mathrm{d}}}

\newcommand{\T}{\ensuremath{\mathrm{\scriptscriptstyle T}}}
\usepackage{bm}

\newcommand{\betab}{\ensuremath{\bm\beta}}

\newcommand{\etab}{\ensuremath{\bm\eta}}
\newcommand{\thetab}{\ensuremath{\bm\theta}}

\usepackage{amsmath}

\newcommand{\dto}{\ensuremath{\overset{\mathrm{d}}{\rightarrow}}}

\usepackage{mathrsfs}

\newcommand{\R}{R}

\newcommand{\subscript}[2]{$#1 _ #2$}
\renewcommand{\baselinestretch}{1.5}

\newcommand{\blind}{1}
  \date{}

\begin{document}

\def\spacingset#1{\renewcommand{\baselinestretch}%
{#1}\small\normalsize} \spacingset{1}


\if1\blind
{
\title{Regression Type Models for Extremal Dependence}
\author{Linda Mhalla$^\text{a}$, Miguel~de Carvalho$^\text{b}$, Val\'erie Chavez-Demoulin$^\text{c,}$\thanks{\textsc{contact}~~~Val\'erie Chavez-Demoulin
    (\href{mailto:valerie.chavez@unil.ch}{\textit{valerie.chavez@unil.ch}}), Faculty of Business and Economics (HEC), Universit\'e de Lausanne, Switzerland. \\ Supplementary materials for this article are available online.} \\
\emph{$^a$Geneva School of Economics and Management (GSEM), Universit\'e de Gen\`eve, Switzerland; $^b$School of Mathematics, University of Edinburgh,  UK; $^c$Faculty of Business and Economics (HEC), Universit\'e de Lausanne, Switzerland.}}

  \maketitle
} \fi

\if0\blind
{
  \bigskip
  \bigskip
  \bigskip
  \begin{center}
    {\Large\bf Regression Type Models for Extremal Dependence}
\end{center}
  \medskip
} \fi

\bigskip
\begin{abstract}
We propose a vector generalized additive modeling framework for taking into account the effect of covariates on angular density functions in a multivariate extreme value context. The proposed methods are tailored for settings where the dependence between extreme values may change according to covariates. We devise a maximum penalized log-likelihood estimator, discuss details of the estimation procedure, and derive its consistency and asymptotic normality. The simulation study suggests that the proposed methods perform well in a wealth of simulation scenarios by accurately recovering the true covariate-adjusted angular density. Our empirical analysis reveals relevant dynamics of the dependence between extreme air temperatures in two alpine resorts during the winter season. Supplementary materials for this article are available online.
\end{abstract}

\noindent%
{\textsc{keywords}:} Angular density; Covariate-adjustment; Penalized log-likelihood; Statistics of multivariate extremes; VGAM.
\vfill

\newpage
\spacingset{1.4} 

\section{\textsf{Introduction}}
\label{sec:intro}


In this paper, we address an extension of the standard approach for modeling non-stationary univariate extremes to the multivariate setting. In the univariate context, the limiting distribution for the maximum of a sequence of independent and identically distributed random variables, derived by \cite{fisher1928}, is given by a generalized extreme value distribution characterized by three parameters: $\mu$ (location), $\sigma$ (scale), and $\xi$ (shape). To take into account the effect of a vector of covariates $\mathbf{x}$, one can let these parameters depend on $\mathbf{x}$, and the resulting generalized extreme value distribution takes the form
\begin{equation}
  G_{(\mu_\mathbf{x}, \sigma_\mathbf{x}, \xi_\mathbf{x})}(y) = \exp\bigg[-\bigg\{1 + \xi_\mathbf{x} \bigg(\frac{y-\mu_\mathbf{x}}{\sigma_\mathbf{x}}\bigg)\bigg\}_{+}^{-1/\xi_\mathbf{x}}\bigg],
  \label{nonsta}
\end{equation}
where $(a)_{+} = \max\{0, a\}$; see \citet[][ch.~6]{coles2001}, \cite{Pauli2001}, \cite{chavez_davison}, \cite{Yee2007}, \cite{wang2009}, \cite{eastoe2009}, and \cite{chavez_davison} for related approaches.

In the multivariate context, consider $\mathbf{Y}^i=\left(Y^i_{1},\ldots,Y^i_{d} \right)^\T$ independent and identically distributed random vectors with joint distribution $F$, and unit Fr\'echet marginal distribution functions $F_j(y)=\exp(-1/y)$, for $y > 0$. Pickands' representation theorem \citep[][Theorem~8.1]{coles2001} states that the law of the standardized componentwise maxima, $\mathbf{M}_n= n^{-1}\max\{\mathbf{Y}^1, \ldots, \mathbf{Y}^n\}$, converges in distribution to a multivariate extreme value distribution, $G_H(\mathbf{y}) = \exp \left\lbrace -V_H(\mathbf{y}) \right\rbrace,$ with 
\begin{equation}
  V_H(\mathbf{y}) = \int_{S_d} \max 
  \bigg(\frac{w_1}{y_1}, \ldots, \frac{w_d}{y_d}\bigg) \,\dif H(\mathbf{w}).
 \label{exponent_V}
\end{equation}
Here $H$ is the so-called angular measure, that is, a positive finite measure on the unit simplex $S_d = \left\lbrace (w_1,\ldots,w_d) \in [0, \infty)^{d}: w_1+\cdots+w_d =1 \right\rbrace$ that needs to obey 
\begin{equation}
  \int_{S_d} w_j \, \dif H(\mathbf{w}) = 1, \quad j=1,\dots,d. \label{condition_H_stationary}
\end{equation}
The function $V(\mathbf{y}) \equiv V_H(\mathbf{y})$, is the so-called exponent measure and is continuous, convex, and homogeneous of order $-1$, i.e., $V(t \mathbf{y}) = t^{-1} V(\mathbf{y})$ for all $t>0$.

The class of limiting distributions of multivariate extreme values yields an infinite number of possible parametric representations \citep[ch.~8]{coles2001}, as the validity of a multivariate extreme value distribution is conditional on its angular measure $H$ satisfying the moment constraint \eqref{condition_H_stationary}. Therefore, most literature has focused on the estimation of the extremal dependence structures described by spectral measures or equivalently angular densities \citep{boldi2007, einmahl2009, decarvalho2013, sabourin2014, hanson2017}. Related quantities, such as the Pickands dependence function \citep{Pickands_1981} and the stable tail dependence function \citep{Huang1992,Drees1998}, were investigated by many authors \citep{einmahl2006,Gudendorf_2012,wadsworth2013,marcon2016}. A wide variety of parametric models for the spectral density that allow flexible dependence structures were proposed \citep[sec.~3.4]{kotz2000extreme}.

However, few papers were able to satisfactorily address the challenging but incredibly relevant setting of modeling nonstationarity at joint extreme levels. Some exceptions include \cite{decarvalho2014a}, who proposed a nonparametric approach, where a family of spectral densities is constructed using exponential tilting. \cite{castro2017} developed an extension of this approach based on covariate-varying spectral densities. However, these approaches are limited to replicated one-way ANOVA types of settings. \cite{decarvalho2016b} advocated the use of covariate-adjusted angular densities, and \cite{GuillouEscobar2016} discussed estimation---in the bivariate and covariate-dependent framework---of the Pickands dependence function based on local estimation with a minimum density power divergence criterion. Finally, \cite{Mhalla2017} constructed, in a nonparametric framework, smooth models for predictor-dependent Pickands dependence functions based on generalized additive models. 

Our approach is based on a non-linear model for covariate-varying extremal dependences. Specifically, we develop a vector generalized additive model that flexibly allows the extremal dependence to change with a set of covariates, but---keeping in mind that extreme values are scarce---it borrows strength from a parametric assumption. In other words, the goal is to develop a regression model for the extremal dependence through the parametric specification of 
an extremal dependence structure and then to model the parameters of that structure through a vector generalized additive model (VGAM) \citep{Yee_1996, Yee_2015}. One major advantage over existing methods is that our model may be used for handling an arbitrary number of dimensions and covariates of different types, and it is straightforward to implement, as illustrated in the \R~code \citep{rdevelopmentcoreteam2016} in the Supplementary Materials.

The remainder of this paper is organized as follows. In Section~\ref{sec:methods} we introduce the proposed model for covariate-adjusted extremal dependences. In Section~\ref{sec:estimation} we develop our penalized likelihood approach and give details on the asymptotic properties of our estimator. In Section~\ref{subsec:simulation} we assess the performance of the proposed methods. An application to extreme temperatures in the Swiss Alps is given in Section~\ref{sec:application}. We close the paper in Section~\ref{sec:discuss} with a discussion. 

\section{\textsf{Flexible Covariate-Adjusted Angular Densities}} \label{sec:methods}

\subsection{\textsf{Statistics of Multivariate Extremes: Preparations and Background}}\label{sec:preps}
The functions $H$ and $V$ in \eqref{exponent_V} can be used to describe the structure of dependence between the extremes, as in the case of independence between the extremes, where $V(\mathbf{y}) = \sum_{j = 1}^d 1/y_j$, and in the case of perfect extremal dependence, where $V(\mathbf{y}) = \max\{1/y_1, \ldots, 1/y_d\}$. As a consequence, if $H$ is differentiable with angular density denoted $h$, the more mass around the barycenter of $S_d$, $(d^{-1}, \ldots, d^{-1})$, the higher the level of extremal dependence. Further insight into these measures may be obtained by considering the point process $P_n = \{ n^{-1}\mathbf{Y}^i: i=1,\ldots,n \}$. Following \cite{deHaan1977} and \citet[sec.~5.3]{Resnick1987}, as $n \rightarrow \infty$, $P_n$ converges to a non-homogeneous Poisson point process $P$ defined on $[\mathbf{0}, \boldsymbol{\infty}) \setminus \{\mathbf{0}\}$ with a mean measure $\mu$ that verifies
\begin{equation}
\mu(A_{\mathbf{y}}) = V(\mathbf{y}), \notag
\end{equation}
where $A_{\mathbf{y}}= \mathbb{R}^d \setminus \left( \left[ -\mathbf{\infty},y_1\right] \times \cdots \times \left[ -\mathbf{\infty},y_d\right]\right)$. 

There are two representations of the intensity measure of the limiting Poisson point process $P$ that will be handy for our purposes. First, it holds that
\begin{equation}
\mu(\dif \mathbf{y}) = - V_{1:d}(\mathbf{y}) \ \dif \mathbf{y}, \label{mu_V}
\end{equation}
with $V_{1:d}$ being the derivative of $V$ with respect to all its arguments \citep[sec.~5.4]{Resnick1987}. Second, another useful factorization of the intensity measure $\mu(\dif \mathbf{y})$, called the spectral decomposition, can be obtained using the following decomposition of the random variable $\mathbf{Y} = (Y_{1}, \ldots, Y_d)^{\T}$ into radial and angular coordinates,
\begin{equation}
(R, \mathbf{W}) = \left(\norm{\mathbf{Y}}, \dfrac{\mathbf{Y}}{\norm{\mathbf{Y}}}\right),
\label{spectral_decomposition}
\end{equation}
where $\norm{\cdot}$ denotes the $L_1$-norm. Indeed, it can be shown that \citep[sec.~8.2.3]{Beirlant_2004} the limiting intensity measure factorizes across radial and angular components as follows:  
\begin{equation*}
\mu(\dif \mathbf{y}) = \mu(\dif r \times \, \dif \mathbf{w} ) =  \frac{\dif r}{r^2}\, \dif H(\mathbf{w}).
\end{equation*}
The spectral decomposition \eqref{spectral_decomposition} allows the separation of the marginal and the dependence parts in the multivariate extreme value distribution $G_H$, with the margins being unit Fr\'{e}chet and the dependence structure being described by the angular measure $H$. 

The inference approach that we build on in this paper was developed by \cite{Coles_1991} and is based on threshold excesses; see \cite{huser_2014} for a detailed review of likelihood estimators for multivariate extremes. The set of extreme events is defined as the set of observations with radial components exceeding a high fixed threshold, that is, the observations belonging to the extreme set, 
\begin{equation*}
E_{\mathbf{r}}= \bigg\{(y_1, \ldots, y_d) \in (0, \infty)^d: \sum_{j=1}^{d} \frac{y_j}{r_j} > 1\bigg\},
\end{equation*}
with $\mathbf{r} = (r_1, \ldots, r_d)$ being a large threshold vector. Since the points $n^{-1} \mathbf{Y}^i$ are mapped to the origin for non-extreme observations, the threshold $\mathbf{r}$ needs to be sufficiently large for the Poisson approximation to hold. Note that, $\mathbf{Y}^i \in E_{\mathbf{r}}$, if and only if, 
\begin{equation}
R_i = \norm{\mathbf{Y}^i} > \left( \sum_{j=1}^{d} \dfrac{W_{i, j}}{r_j}\right) ^{-1}, \quad \text{where } W_{i, j} = \dfrac{Y^i_{j}}{R_i}. \notag
\end{equation}
Hence, the expected number of points of the Poisson process $P$ located in the extreme region $E_{\mathbf{r}}$ is
\begin{eqnarray}
\mu(E_{\mathbf{r}}) &=& \int_{S_d} \int_{\left( \sum_{j=1}^{d} \frac{w_{j}}{r_j}\right) ^{-1}}^{\infty} \dfrac{\dif r}{r^2} \, \dif H(\mathbf{w}) \notag \\
&=& \int_{S_d} \left( \sum_{j=1}^{d} \dfrac{w_{j}}{r_j}\right) \, \dif H(\mathbf{w}) \notag \\
&=& \sum_{j=1}^d \frac{1}{r_j} \int_{S_d} w_j \, \dif H(\mathbf{w}) = \sum_{j=1}^{d} \frac{1}{r_j}. \label{mu_A}
\end{eqnarray}
Now, we can explicitly formulate the Poisson log-likelihood over the set $E_{r}$,
\begin{equation}
\ell_{E_{\mathbf{r}}} (\thetab) =  - \mu(E_{\mathbf{r}}) +  \sum_{i=1}^{n_\mathbf{r}} \log \left\lbrace \mu(\dif R_i \times \dif \mathbf{W}_i)\right\rbrace, \label{likelihood_A}
\end{equation}
where $\thetab$ represents the $p-$vector of parameters of the measure $\mu$ and $n_\mathbf{r}$ represents the number of reindexed observations in the extreme set $E_{\mathbf{r}}$. Using \eqref{mu_A}, the first term in \eqref{likelihood_A} can be omitted when maximizing the Poisson log-likelihood, which, using \eqref{mu_V}, boils down to
\begin{equation}
\ell_{E_{\mathbf{r}}} (\thetab) \equiv \sum_{i=1}^{n_\mathbf{r}} \log \left\lbrace -V_{1:d}(\mathbf{Y}^i ; \thetab)\right\rbrace. \label{likelihood_V}
\end{equation}
Thanks to the differentiability of the exponent measure $V$ and the support of the angular measure $H$ in the unit simplex $S_d$, we can use the result of \citet[][Theorem~1]{Coles_1991} that relates the angular density to the exponent measure via
\begin{equation}
V_{1:d}(\mathbf{y}; \thetab) = - \|\mathbf{y}\|^{-(d+1)} h\left( \dfrac{y_{1}}{\|\mathbf{y}\|}, \ldots, \dfrac{y_{d}}{\|\mathbf{y}\|} ; \thetab \right) \notag
\end{equation}
and reformulate the log-likelihood \eqref{likelihood_V} as follows
\begin{eqnarray}
\ell_{E_{\mathbf{r}}} (\thetab) &\equiv& -(d+1) \sum_{i=1}^{n_\mathbf{r}} \log  \norm{\mathbf{Y}^{i}}  + \sum_{i=1}^{n_\mathbf{r}} \log \left\lbrace h \left( \dfrac{Y^i_{1}}{\norm{\mathbf{Y}^{i}}}, \ldots, \dfrac{Y^i_{d}}{\norm{\mathbf{Y}^{i}}}; \thetab \right)\right\rbrace  \label{likelihood_h} \\
&=& \sum_{i=1}^{n_\mathbf{r}} \ell_{E_{\mathbf{r}}} (\mathbf{Y}^i,\thetab). \notag
\end{eqnarray}

\subsection{\textsf{Vector Generalized Additive Models for Covariate-Adjusted Angular Densities}}
\label{sec:gamspectral}
Our starting point for modeling is an extension of \eqref{nonsta} to the multivariate setting. Whereas the model in~\eqref{nonsta} is based on indexing the parameters of the univariate extreme value distribution with a regressor, here we index the parameter ($H$) of a multivariate extreme value distribution ($G_H$) with a regressor $\mathbf{x}=(x_1,\ldots, x_q)^{\T} \in \mathcal{X} \subset \mathbb{R}^q$. Our target object of interest is thus given by a family of covariate-adjusted angular measures $H_{\mathbf{x}}$ obeying
\begin{equation}
  \int_{S_d} w_j \dif H_{\mathbf{x}}(\mathbf{w}) = 1, \quad j=1,\dots,d. \notag
\end{equation}
Of particular interest is the setting where $H_{\mathbf{x}}$ is differentiable, in which case the covariate-adjusted angular density can be defined as $h_{\mathbf{x}}(\mathbf{w}) = \dif H_{\mathbf{x}}/\dif\mathbf{w}$. This yields a corresponding family of covariate-indexed multivariate extreme value distributions 
\begin{equation}
  G_{\mathbf{x}}(\mathbf{y}) = \exp\bigg\{- \int_{S_d} \max\bigg(
  \frac{w_1}{y_1}, \ldots ,\frac{w_d}{y_d}\bigg) \, \text{d}H_{\mathbf{x}}(\mathbf{w})\bigg\}.
  \notag
\end{equation}
Other natural objects depending on $G_{\mathbf{x}}$ can be readily defined, such as the covariate-adjusted extremal coefficient, $\vartheta_{\mathbf{x}}$, which solves
\begin{equation}\label{covext}
  G_{\mathbf{x}}(y \boldsymbol{1}_d) = \exp(- \vartheta_{\mathbf{x}} / y), \quad y > 0,
\end{equation}
where $\boldsymbol{1}_d$ is a $d-$vector of ones. Here, $\vartheta_{\mathbf{x}}$ ranges from 1 to $d$, and the closer $\vartheta_{\mathbf{x}}$ is to one, the closer we get to the case of complete dependence at that value of the covariate.

\begin{figure}[!h]
  \centering  
  \textbf{Covariate-adjusted angular densities} \subfloat{\includegraphics[width=0.3\textwidth]{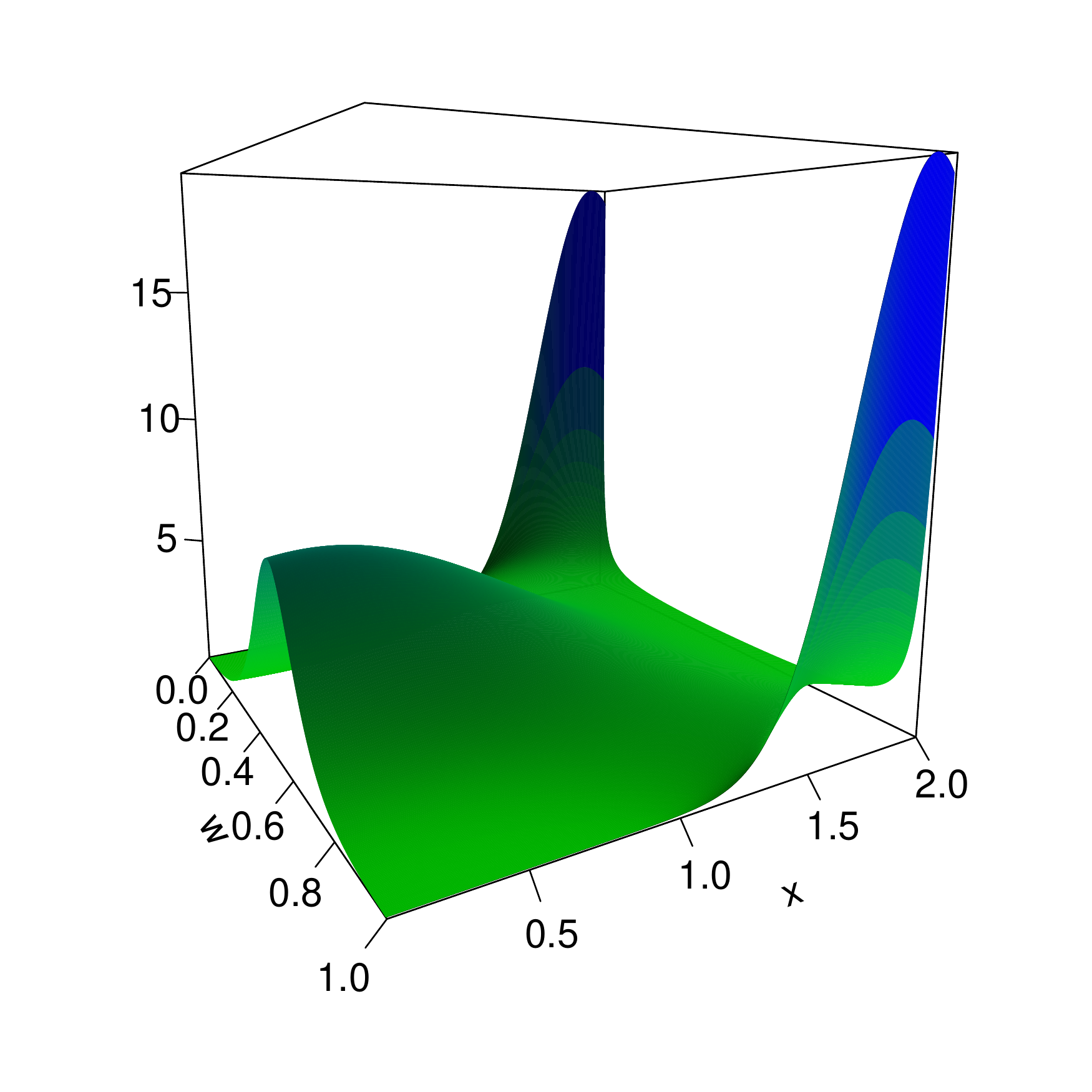}} \subfloat{\includegraphics[width=0.3\textwidth]{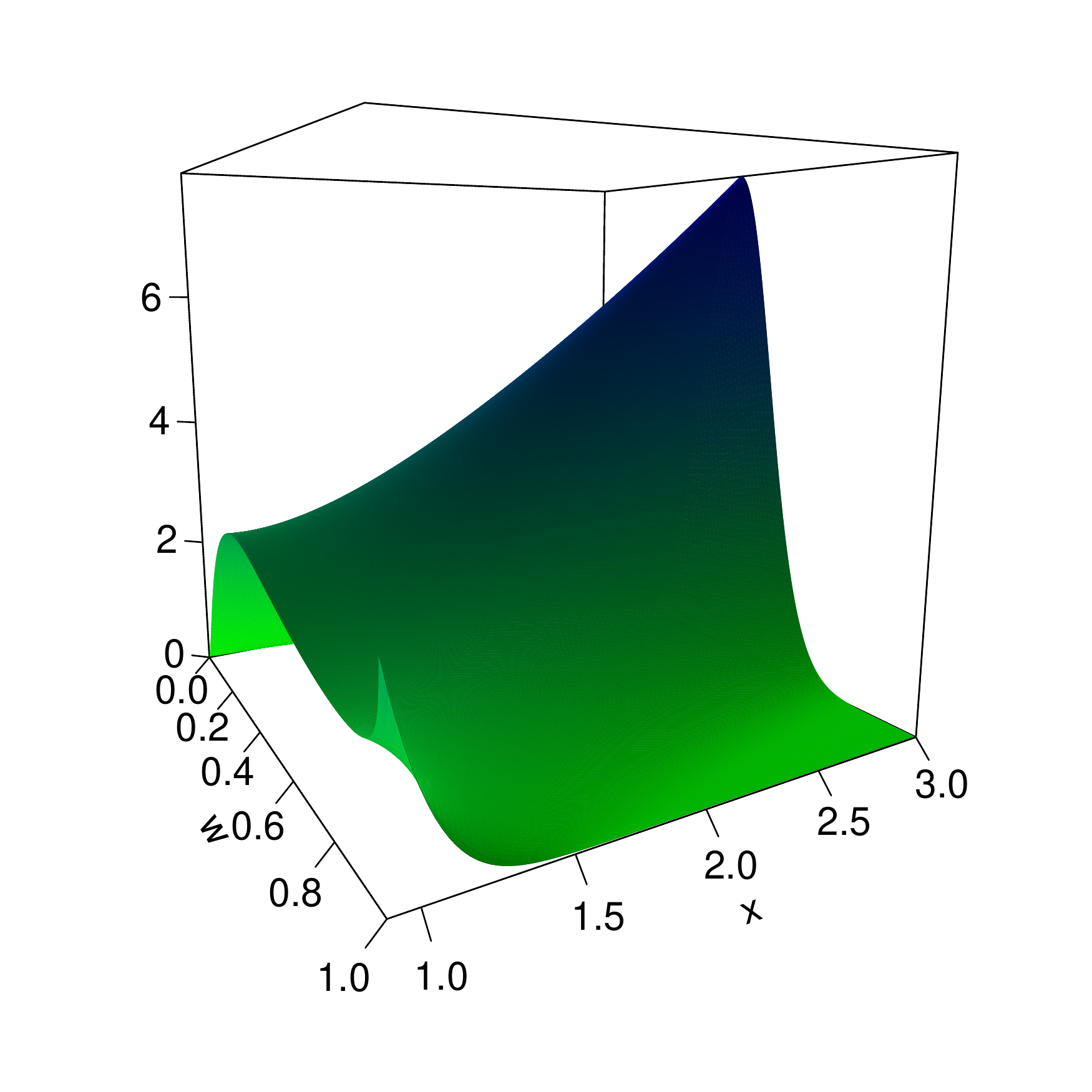}}
  \subfloat{\includegraphics[width=0.3\textwidth]{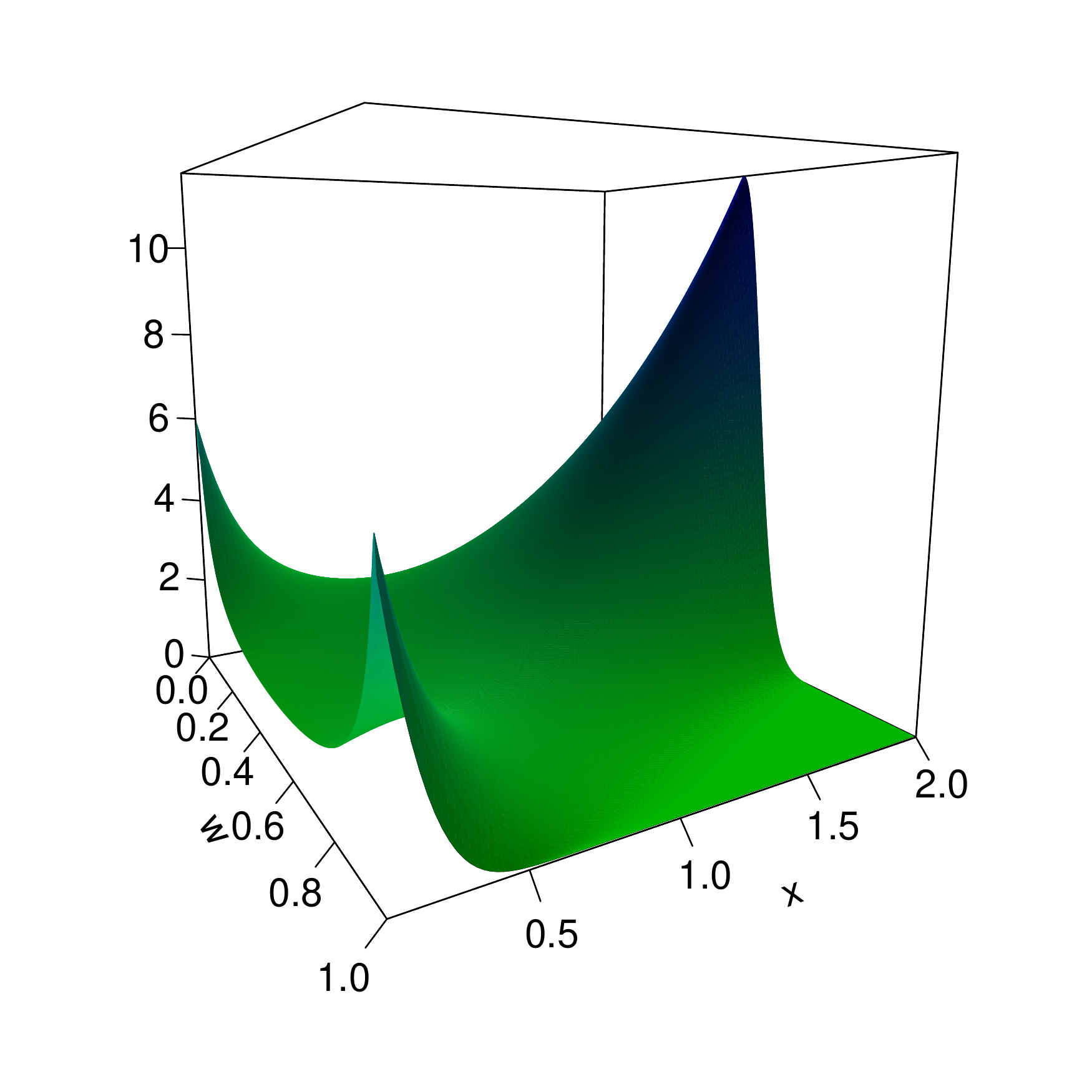}} \\
  \textbf{Covariate-adjusted extremal coefficient} \\
  \subfloat{\includegraphics[width=0.3\textwidth]{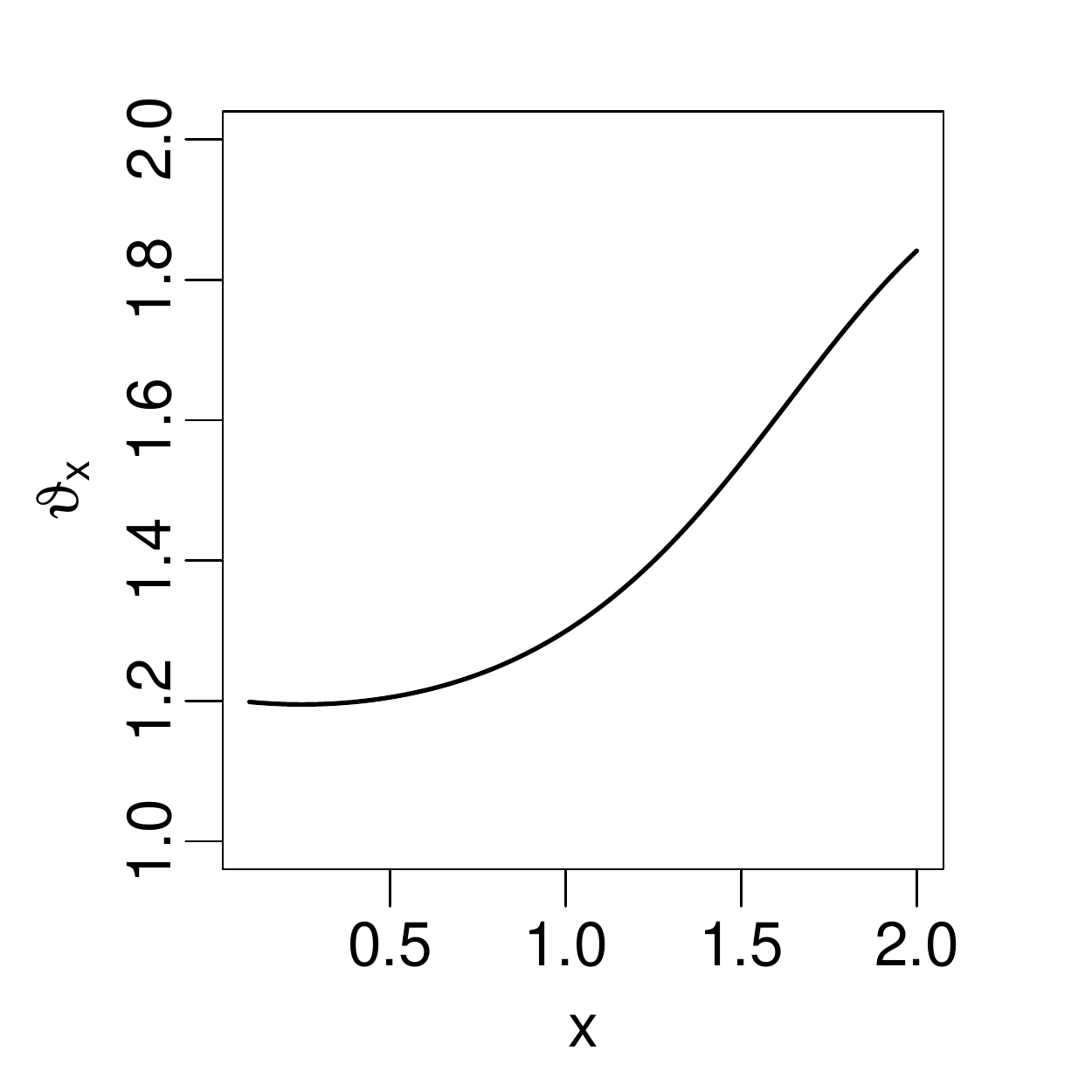}} \subfloat{\includegraphics[width=0.3\textwidth]{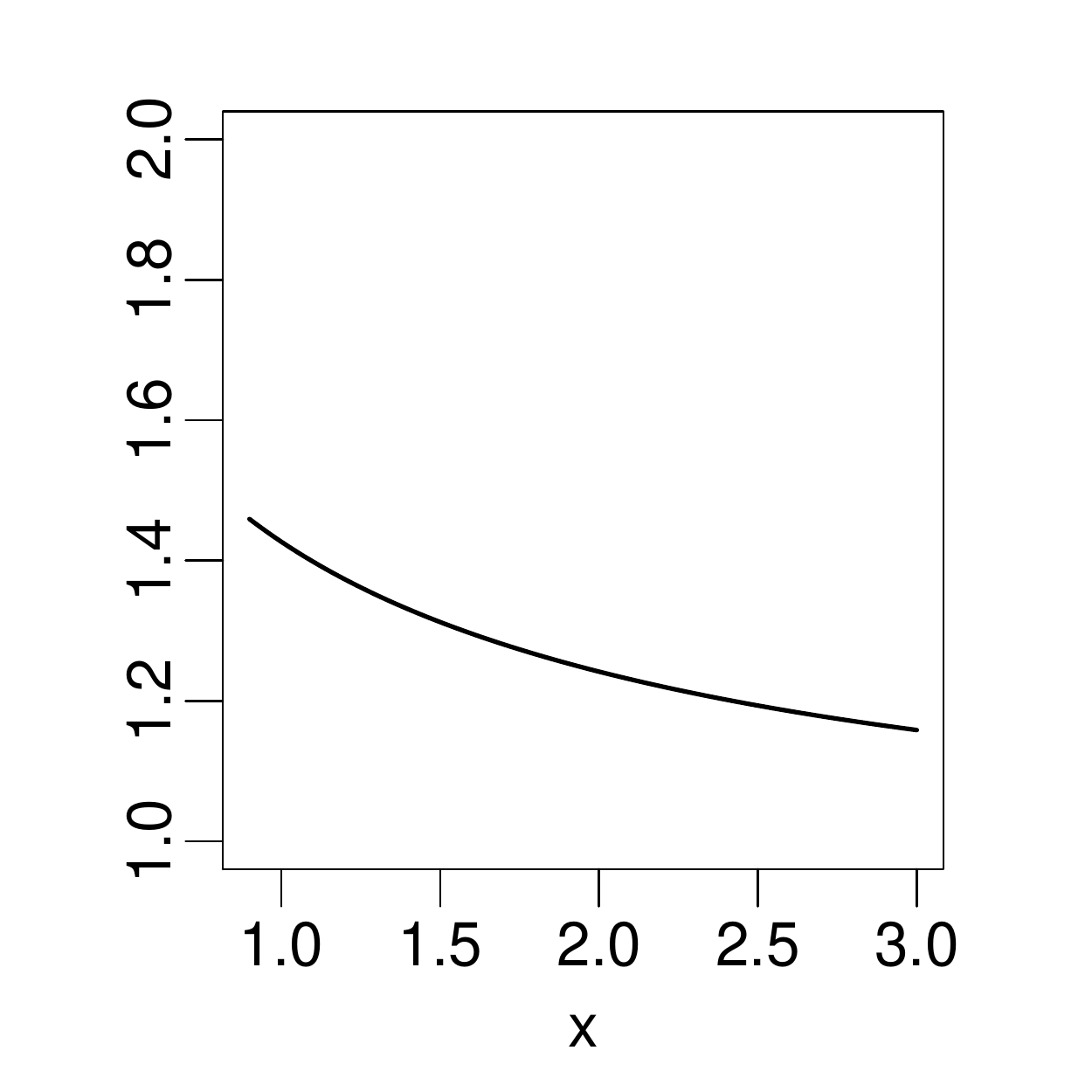}}
  \subfloat{\includegraphics[width=0.3\textwidth]{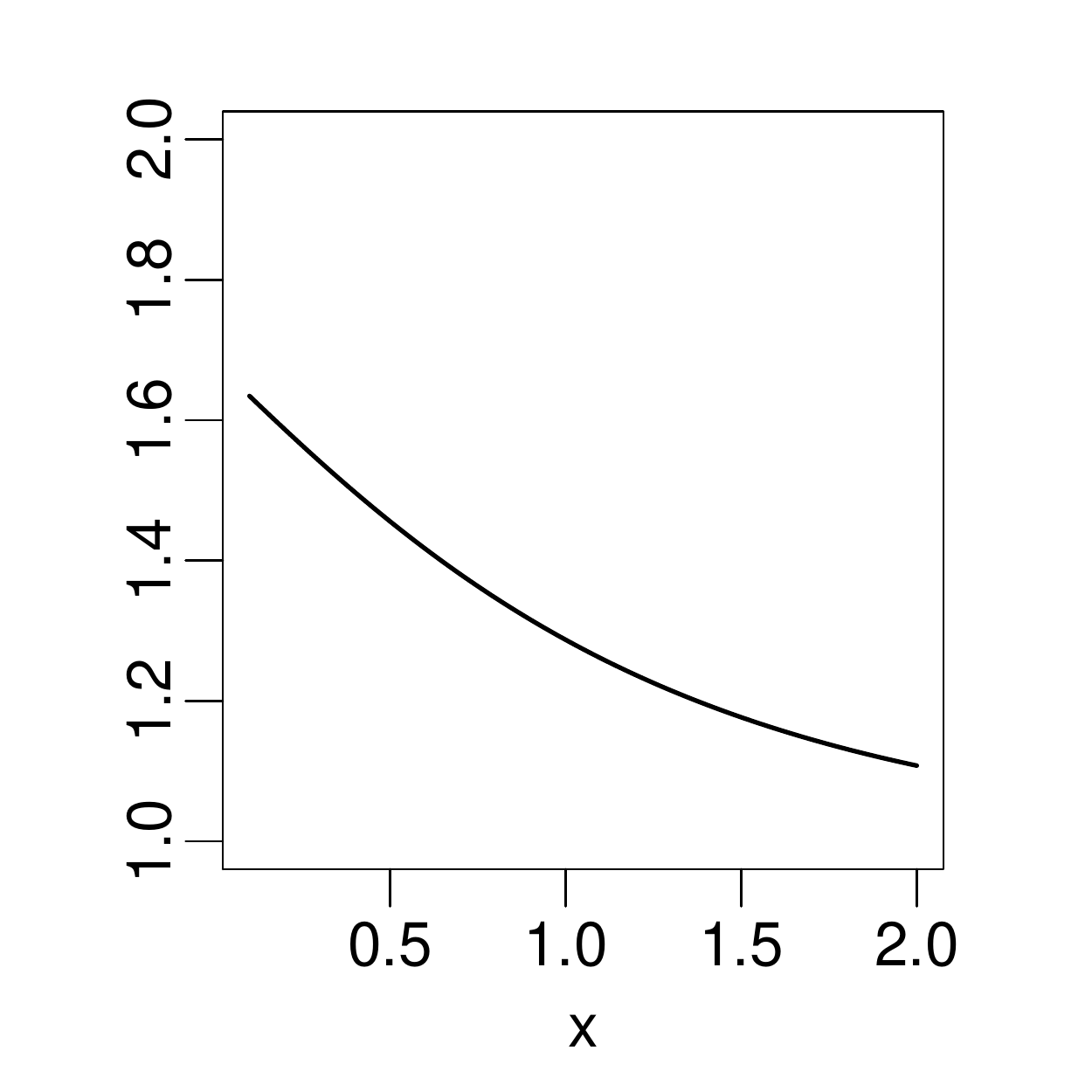}} \\
  \caption{\footnotesize Covariate-adjusted angular densities and extremal coefficients of logistic (left panels), Dirichlet (middle panels), and H\"{u}sler--Reiss (right panels) models, corresponding, respectively, to the specifications in Examples~\ref{ex:logistic_surf}, \ref{ex:Dirichlet_surf}, and \ref{ex:Husler_surf}.}
\label{fig:spectral_surfaces}
\end{figure}

\begin{figure}[!h]
  \centering
  \subfloat{\includegraphics[width=0.3\textwidth]{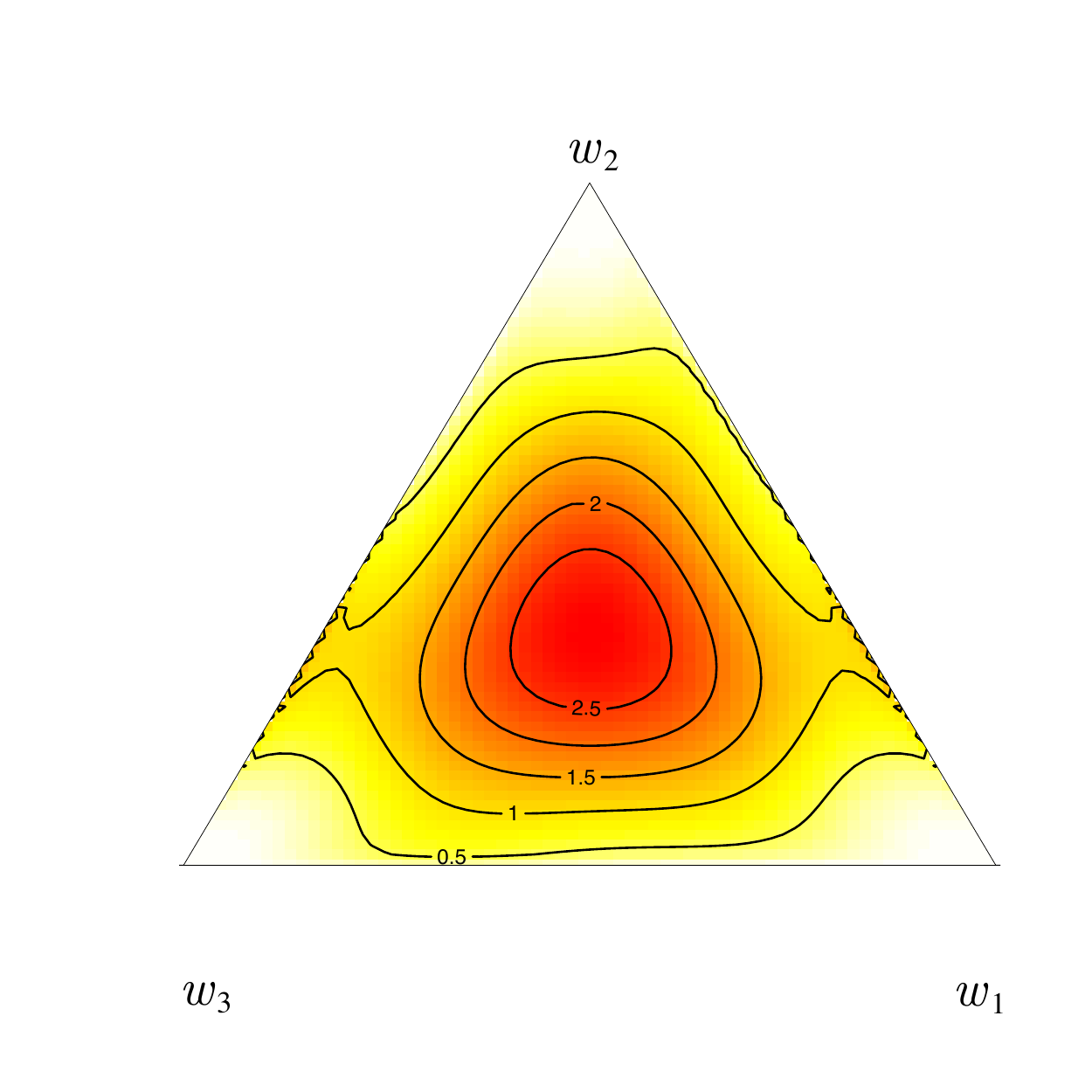}}
  \subfloat{\includegraphics[width=0.3\textwidth]{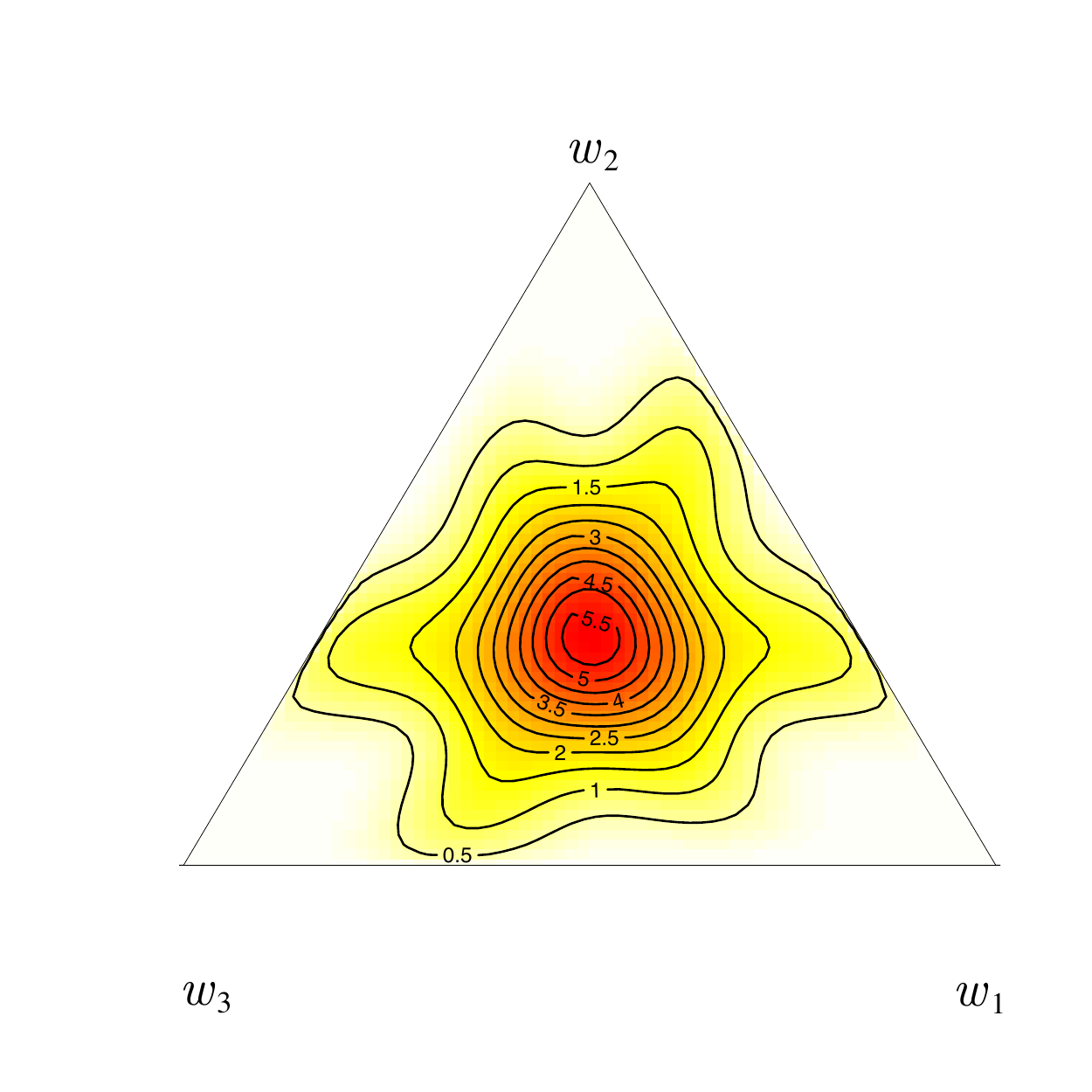}}
  \subfloat{\includegraphics[width=0.3\textwidth]{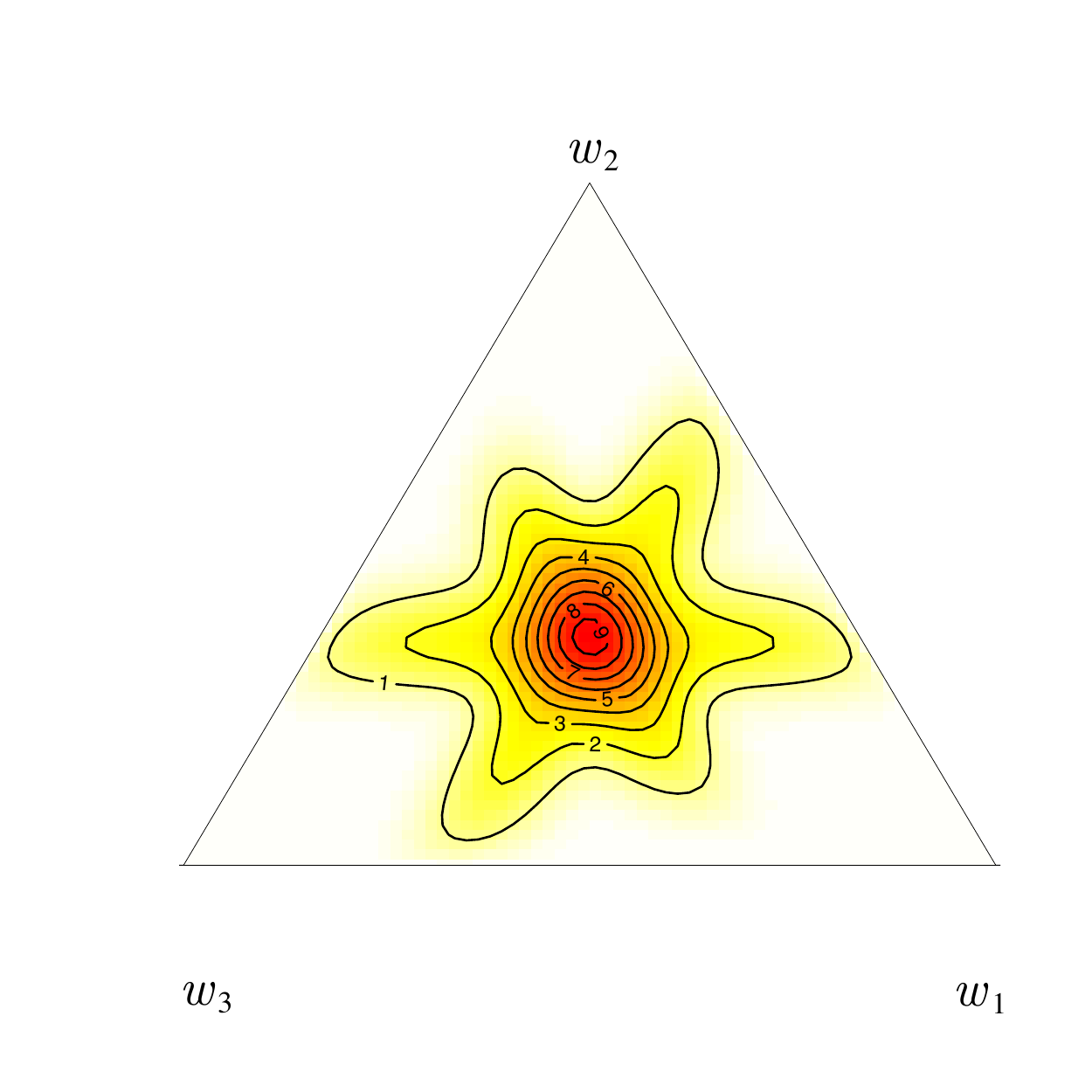}} 
  \caption{\footnotesize Trivariate covariate-adjusted angular density of the pairwise beta model corresponding to the specifications in Example~\ref{ex:pairwise_beta} with $x=1.5$ (left), $x=2.46$ (middle), and $x=3.22$ (right).} 
\label{fig:pBeta_angular_density}
\end{figure}

\noindent Some parametric models \citep{logistic,Coles_1991,HR,pairwise.beta.Cooley} are used below to illustrate the concept of covariate-adjusted angular densities and of a covariate-adjusted extremal coefficient, and we focus on the bivariate and trivariate settings for the sake of illustrating ideas. To develop insight and intuition on these models, see Figures~\ref{fig:spectral_surfaces} and \ref{fig:pBeta_angular_density}.

\begin{example}[Logistic angular surface]\normalfont
Let
\begin{equation}
h_{\mathbf{x}}(w) = (1/\alpha_{\mathbf{x}} - 1)  \left\lbrace w (1-w)\right\rbrace ^{-1-1/\alpha_{\mathbf{x}}} \{w^{-1/\alpha_{\mathbf{x}}} + (1-w) ^{-1/\alpha_{\mathbf{x}}}\}  ^{\alpha_{\mathbf{x}}-2}, \quad w \in (0,1),
\notag
\end{equation}
with $\alpha: \mathcal{X} \subset \mathbb{R}^q \to (0, 1]$. In Figure~\ref{fig:spectral_surfaces}~(left) we represent the case $\alpha_x= \exp\{\eta(x)\}/[1+\exp\{\eta(x)\}]$, with $\eta(x)=x^2-0.5x-1$ and $x \in \mathcal{X} = [0.1,2]$. This setup corresponds to be transitioning between a case of relatively high extremal dependence (lower values of $x$) to a case where we approach asymptotic independence (higher values of $x$). 
\label{ex:logistic_surf}
\end{example}

\begin{example}[Dirichlet angular surface]\normalfont
Let
\begin{equation}
h_{\mathbf{x}}(w) = \frac{\alpha_{\mathbf{x}} \beta_{\mathbf{x}} \Gamma(\alpha_{\mathbf{x}} +  \beta_{\mathbf{x}} + 1) (\alpha_{\mathbf{x}} w)^{\alpha_{\mathbf{x}} - 1} \{\beta_{\mathbf{x}}(1 - w)\}^{\beta_{\mathbf{x}} - 1}}{ \Gamma(\alpha_{\mathbf{x}}) \Gamma(\beta_{\mathbf{x}}) \{\alpha_{\mathbf{x}} w + \beta_{\mathbf{x}} (1 - w)\}^{\alpha_{\mathbf{x}} + \beta_{\mathbf{x}} + 1}}, \quad w \in (0,1),
\notag
\end{equation}
with $\alpha: \mathcal{X} \subset \mathbb{R}^q \to (0, \infty)$ and $\beta: \mathcal{X} \subset \mathbb{R}^q \to (0, \infty)$. In Figure~\ref{fig:spectral_surfaces}~(middle) we consider the case $\alpha_x=\exp(x)$ and $\beta_x=x^2$, with $x \in [0.9,3]$. Note the different schemes of extremal dependence induced by the different values of the covariate $x$ as well as the asymmetry of the angular surface underlying this model. 
\label{ex:Dirichlet_surf}
\end{example}

\begin{example}[H\"{u}sler--Reiss angular surface]\normalfont
Let
\begin{equation}
h_{\mathbf{x}}(w) = \frac{\lambda_{\mathbf{x}}}{ w (1 - w)^2 (2 \pi)^{1 / 2}} 
\exp\bigg\{- \dfrac{\left[  2 + \lambda_{\mathbf{x}}^2 \log\left\lbrace w/(1 - w) \right\rbrace  \right]^2}{8 \lambda_{\mathbf{x}}^2}\bigg\}, \quad w \in (0, 1),
\notag
\end{equation}
where $\lambda: \mathcal{X} \subset \mathbb{R}^{q} \to (0, \infty)$. In Figure~\ref{fig:spectral_surfaces}~(right) we consider the case $\lambda_{x} = \exp(x)$, with $x \in [0.1,2]$. Under this specification, lower values of $x$ correspond to lower levels of extremal dependence, whereas higher values of $x$ correspond to higher levels of extremal dependence.
\label{ex:Husler_surf}
\end{example}

\begin{example}[Pairwise beta angular surface]\normalfont
Let
\begin{eqnarray}
h_{\mathbf{x}}(\mathbf{w}) &=& \dfrac{\Gamma(3\alpha_{\mathbf{x}}+1)}{\Gamma(2\alpha_{\mathbf{x}}+1) \Gamma(\alpha_{\mathbf{x}})} \sum_{1 \leq i<j \leq 3} h_{i,j_{\mathbf{x}}}(\mathbf{w}), \quad \notag \\
h_{i,j_{\mathbf{x}}}(\mathbf{w})&=& (w_i+w_j)^{2\alpha_{\mathbf{x}}-1} \left\lbrace 1-(w_i+w_j)\right\rbrace ^{\alpha_{\mathbf{x}}-1} \dfrac{\Gamma(2\beta_{i,j_{\mathbf{x}}})}{\Gamma^2(\beta_{i,j_{\mathbf{x}}})} \left( \dfrac{w_i}{w_i+w_j}\right)^{\beta_{i,j_{\mathbf{x}}}-1}  \left( \dfrac{w_j}{w_i+w_j}\right)^{\beta_{i,j_{\mathbf{x}}}-1}, \quad \notag
\end{eqnarray}
where $\mathbf{w}=(w_1,w_2,w_3) \in S_3$ and $\alpha,\beta_{i,j}: \mathcal{X} \subset \mathbb{R}^{q} \to (0, \infty)$ for $1\leq i<j \leq 3$. In Figure~\ref{fig:pBeta_angular_density}, we consider the case $\alpha_{\mathbf{x}}=\exp\{\exp(x)\}$, $\beta_{1,2_{\mathbf{x}}}=\exp(x)$, $\beta_{1,3_{\mathbf{x}}}=x+1$, and $\beta_{2,3_{\mathbf{x}}}=x+2$, with $x \in [0.8,3.3]$. For the different considered values of $x$, different strengths of global and pairwise dependences can be observed. The mass is concentrated mostly at the center of the simplex due to a large global dependence parameter $\alpha_{\mathbf{x}}$, compared to the pairwise dependence parameters.
\label{ex:pairwise_beta}
\end{example}

The previous parametric models provide some examples of covariate-adjusted angular surfaces $h_{\mathbf{x}}$. But, how can we learn about $h_{\mathbf{x}}$ from the data? Suppose we observe the regression data $\{(\mathbf{x}^i, \mathbf{Y}^i)\}_{i = 1}^n$, with $(\mathbf{x}^i, \mathbf{Y}^i) \in \mathcal{X} \times \mathbb{R}^d$, and where we assume that $\mathbf{Y}^i=\left(Y^i_{1},\ldots,Y^i_{d} \right)^\T$ are independent random vectors with unit Fr\'echet marginal distributions. Using a similar approach as in Section~\ref{sec:preps}, we convert the raw sample into a pseudo-sample of cardinality $n_{\mathbf{r}}$,
\begin{equation*}
  \{(\mathbf{x}^i, \mathbf{Y}^i): \mathbf{Y}^i \in E_{\mathbf{r}}\},
\end{equation*}
and use the latter reindexed data to learn about $h_{\mathbf{x}}$. \\
Without loss of generality, we restrain ourselves to the bivariate extreme value framework ($d=2$), so that 
\begin{equation}
h_{\mathbf{x}}\left(\dfrac{Y^i_{1}}{\|\mathbf{Y^i}\|},\dfrac{Y^i_{2}}{\|\mathbf{Y^i}\|}\right) = h_{\mathbf{x}}\left(w_i,1-w_i\right) \equiv  h_{\mathbf{x}}(w_i) \notag, \quad \text{for } w_i \in [0,1], \quad i=1,\ldots, n_{\mathbf{r}},
\end{equation}
that is, the dimension of the angular observations $w_i$ is $M=d-1=1$. We model $h_{\mathbf{x}}(\cdot)$ using $h(\cdot; \thetab_{\mathbf{x}})$, where the parameter underlying the dependence structure
\begin{eqnarray}
 \thetab_{\mathbf{x}} &=& (\theta_{1\mathbf{x}^1}, \dots, \theta_{1\mathbf{x}^{n_{\mathbf{r}}}}, \dots, \theta_{p\mathbf{x}^1}, \ldots \theta_{p\mathbf{x}^{n_{\mathbf{r}}}})^{\T} \in \mathbb{R}^{p n_{\mathbf{r}}}, \notag \\
 \quad \mathbf{x} &=& (\mathbf{x}^1,\ldots, \mathbf{x}^{n_{\mathbf{r}}})^{\T} \in \mathcal{X}^{n_{\mathbf{r}}}=\left( \mathcal{X}_1 \times \cdots \times \mathcal{X}_q\right) ^{n_{\mathbf{r}}}  \subseteq \mathbb{R}^{q n_{\mathbf{r}}}
 \notag
\end{eqnarray}
is specified through a vector generalized additive model (VGAM) \citep{Yee_1996}. Specifically, we model $h_{\mathbf{x}}(w)$ using a fixed family of parametric extremal dependence structures $h(w; \thetab_{\mathbf{x}})$ with a covariate-dependent set of parameters $\thetab_{\mathbf{x}}$. To learn about $\thetab_{\mathbf{x}}$ from the pseudo-sample, we use a vector generalized additive model, which takes the form
\begin{equation}
\etab(\mathbf{x}) \equiv \etab = \mathbf{H}_0 \betab_{[0]} + \sum_{k=1}^q \mathbf{H}_k \mathbf{f}_{k}(\mathbf{x}_k). \label{gam_theta}
\end{equation}
Here,
\begin{itemize}
\item $\etab = \mathbf{g}\left(\thetab_{\mathbf{x}}\right) = \left( g_1(\theta_{1x^1}), \ldots, g_1(\theta_{1x^{n_{\mathbf{r}}}}), \ldots, g_p(\theta_{px^1}), \ldots, g_p(\theta_{px^{n_{\mathbf{r}}}})\right) ^\T$ is the vector of predictors and $g_l$ is a link function that ensures that $\theta_{l\cdot}$ is well defined, for $l=1,\ldots,p$,
\item $\betab_{[0]}$ is a $p n_{\mathbf{r}}-$vector of intercepts, with $p$ distinct values each repeated $n_{\mathbf{r}}$ times,
\item $\mathbf{x}_k=\left( x_k^1,\ldots,x_k^{n_{\mathbf{r}}}\right) ^\T \in \mathcal{X}_k^{{n_{\mathbf{r}}}}$, for $k=1,\ldots,q$,
\item $\mathbf{f}_{k}=(\mathbf{f}_{k,1},\ldots,\mathbf{f}_{k,p})^\T$, where $\mathbf{f}_{k,l}=(f_{k,l}(x^1_{k}), \ldots,f_{k,l}(x^{n_{\mathbf{r}}}_k))^\T,$ and $f_{k,l}: \mathcal{X}_k \rightarrow \mathbb{R}$ are smooth functions supported on $\mathcal{X}_k$, for $k=1,\ldots,q$ and $l=1,\ldots,p$, and
\item $\mathbf{H}_k$ are $p n_{\mathbf{r}} \times p n_{\mathbf{r}}$ constraint matrices, for $k=0,\ldots,q$.
\end{itemize}
The constraint matrices $\mathbf{H}_k$ are important quantities in the VGAM \eqref{gam_theta} that allow the tuning of the effects of the covariates on each of the $p n_{\mathbf{r}}$ components of $\etab$. For example, in Example~\ref{ex:pairwise_beta}, one might want to impose the same smooth effect of a covariate on each of the $\binom{3}{2}$ pairwise dependence parameters and at the same time restrict the effect of this covariate to be zero on the global dependence parameter. To avoid clutter in the notation, we assume from now on that $\mathbf{H}_k= \mathbf{I}_{p n_{\mathbf{r}} \times p n_{\mathbf{r}}}$, for $k=0,\ldots,q$.

The smooth functions $f_{k,l}$ are written as linear combinations of $B$-spline basis functions
\begin{equation}
f_{k,l}(x^i_k)= \sum_{s=1}^{d_k} \beta_{[kl]_s} B_{s,\tilde{q}}(x^i_k), \quad k=1,\ldots,q, \quad l=1,\ldots,p, \quad i=1,\ldots, n_{\mathbf{r}},\notag
\end{equation}
where $B_{s,\tilde{q}}$ is the $s$th $B$-spline of order $\tilde{q}$ and $d_k=\tilde{q}+m_k$, with $m_k$ the number of internal equidistant knots for $\mathbf{x}_k$ \citep[sec.~2.4.5]{Yee_2015}. To ease the notational burden, we suppose without loss of generality that $d_k \equiv \tilde{d}$, for $k=1,\ldots,q$, and define
\begin{equation}
\betab_{[k]} = \left( \beta_{[k1]_1}, \ldots, \beta_{[k1]_{\tilde{d}}}, \ldots, \beta_{[kp]_1}, \ldots, \beta_{[kp]_{\tilde{d}}}\right) ^\T \in \mathbb{R}^{\tilde{d}p}. \notag
\end{equation}
Therefore, the VGAM \eqref{gam_theta}, with identity constraint matrices $\mathbf{H}_k$, can be written as
\begin{equation}
\etab = \betab_{[0]} + \sum_{k=1}^{q} \mathbf{X}_{[k]} \betab_{[k]} = \mathbf{X}_{\text{VAM}} \betab, \label{VGAM_eta}
\end{equation}
where 
\begin{eqnarray}
\begin{cases}
\betab = \begin{pmatrix}
\betab_{[0]} & \betab_{[1]} & \cdots & \betab_{[q]}
\end{pmatrix}^\T \in \mathbf{B} \subset \mathbb{R}^{p(1+q\tilde{d})},  \notag \\
\mathbf{X}_{\text{VAM}} = \begin{pmatrix}
\boldsymbol{1}_{p n_{\mathbf{r}} \times p} & \mathbf{X}_{[1]} & \cdots & \mathbf{X}_{[q]}
\end{pmatrix} \in \mathbb{R}^{pn_{\mathbf{r}} \times \{p(1+q\tilde{d})\}} \notag
\end{cases}
\end{eqnarray}
for some $pn_{\mathbf{r}} \times \tilde{d}p$ submatrices $\mathbf{X}_{[k]}$, $k=1,\ldots,q$. The vector of parameters to be estimated in the VGAM \eqref{VGAM_eta} is $\betab$.

The specification in \eqref{VGAM_eta} makes it possible to simultaneously fit ordinary Generalized Additive Models \citep{wood_Book_2017} in each component of the vector of parameters $\thetab_{\mathbf{x}}$, hence avoiding any non orthogonality-related issues that could arise if the $p$ components were to be treated separately \citep{chavez_davison}. Finally, if the dimension $M$ of the response vector of angular observations $w_i$ is greater than one ($d>2$), then the vector of predictors $\etab$ will instead be a $Mpn_{\mathbf{r}}-$vector and the dimensions of the related quantities in \eqref{VGAM_eta} will change accordingly.

To give the unfamiliar reader insight on some of the quantities introduced above, we identify these quantities in the examples mentioned previously: 
\begin{itemize}
\item In Examples~\ref{ex:logistic_surf} and \ref{ex:Husler_surf}, $d=2$, $M=1$, $p=1$, $q=1$, and $\mathcal{X}=[0.1,2]$. The difference between the VGAMs modeled in these two examples resides in the form of dependence of $\eta$ on $x$ and the link function $g$. In Example~\ref{ex:logistic_surf}, the parameter $\theta_x \in (0,1]$, $\eta=x^2-0.5x-1$, and the link function $g$ is the logit function, whereas in Example~\ref{ex:Husler_surf} the parameter $\theta_x \in (0,\infty)$, $\eta=x$, and the link function $g$ is the logarithm function.
\item In Example~\ref{ex:Dirichlet_surf}, $d=2$, $M=1$, $p=2$, $q=1$, $\mathcal{X}=[0.9,3]$, and $\etab=(x, x)^\T$. The vector of parameters for the bivariate Dirichlet angular density $\thetab_{x} \in (0,\infty)^2$ and the link functions $g_1$ and $g_2$ are the logarithm and the square root functions, respectively.
\item In Example~\ref{ex:pairwise_beta}, $d=3$, $M=2$, $p=4$, $q=1$, $\mathcal{X}=[0.8,3.3]$, and $\etab= (\exp(x), x, \log(x+1), \log(x+2))^\T$. The vector of parameters for the pairwise beta angular density $\thetab_{x} \in (0,\infty)^4$ and the link function $g_l$ is the logarithm function, for $l=1,\ldots,4$.
\end{itemize}

\section{\textsf{Inference and Asymptotic Properties}}
\label{sec:estimation}
The log-likelihood \eqref{likelihood_h} with a covariate-dependent vector of parameters $\thetab_{\mathbf{x}}$ is now written as
\begin{equation}
\ell(\betab) := \sum_{i=1}^{n_\mathbf{r}} \ell \left( \mathbf{Y}^i ; \betab\right)  = \sum_{i=1}^{n_\mathbf{r}} \ell_{E_{\mathbf{r}}} 
\left[  \mathbf{Y}^i,\mathbf{g}^{-1}\{\etab(\mathbf{x}^i)\}\right] , \notag
\end{equation}
where $\mathbf{g}^{-1}$ is the componentwise inverse of $\mathbf{g}$.

Incorporating a covariate-dependence in the extremal dependence model through a non-linear smooth model adds considerable flexibility in the modeling of the dependence parameter $\thetab_{\mathbf{x}}$. The price to pay for this flexibility is reflected in the estimation procedure. The estimation of $\thetab_{\mathbf{x}}$, hence of $\betab$, is performed by maximizing the penalized log-likelihood
\begin{equation}
\ell(\betab, \gammaV) = \ell(\betab) -  \dfrac{1}{2} \mathbf{J}(\gammaV), \label{penalized_likelihood}
\end{equation}
where the penalty term can be written as
\begin{equation}
\mathbf{J}(\gammaV) = \sum_{k=1}^{q} \betab_{[k]}^\T \left\lbrace \mathbf{P}_k \otimes \text{diag}(\gamma_{(1)k}, \ldots, \gamma_{(p)k})\right\rbrace \betab_{[k]} = \betab^\T \mathbf{P}(\gammaV) \betab, \notag
\end{equation}
with $\mathbf{P}(\gammaV)$ a $p(1+q\tilde{d}) \times p(1+q\tilde{d})$ block matrix with a first $p\times p$ block filled with zeros and $q$ blocks, each formed by a $p\tilde{d}\times p\tilde{d}$ matrix $\mathbf{P}_k$ that depends only on the knots of the $B$-spline functions for the covariate $\mathbf{x}_k$. The matrix $\mathbf{P}(\gammaV)$ can be written as $\mathbf{P}(\gammaV)=\tilde{\mathbf{X}}^\T \tilde{\mathbf{X}}$ for some $p(1+q\tilde{d}) \times p(1+q\tilde{d})$ real matrix $\tilde{\mathbf{X}}$. The vectors $\betab_{[k]}$ are defined in \eqref{VGAM_eta}, and $\gamma_{(l)k}$ are termed the smoothing parameters.

\noindent The penalty term in \eqref{penalized_likelihood} controls the wiggliness and the fidelity to the data of the component functions in \eqref{gam_theta} through the vector $\gammaV$ of the smoothing parameters $\gamma_{(l)k}$ for $l=1,\ldots, p$ and $k=1,\ldots,q$. Larger values of $\gamma_{(l)k}$ lead to smoother effects of the covariate $\mathbf{x}_k$ on the $l$th component of $\etab$. 

The maximization of the penalized log-likelihood \eqref{penalized_likelihood} is based on a Newton--Raphson (N--R) algorithm. At each step of the N--R algorithm, a set of smoothing parameters is proposed by outer iteration \citep{wood_Book_2017}, and a penalized iterative reweighted least squares (PIRLS) algorithm is performed, in an inner iteration, to update the model coefficients estimates. We detail the inner fitting procedure in the following section and the outer iteration in Section~\ref{subsec:selection_gamma}.

\subsection{\textsf{Fitting Algorithm}\label{subsec:fitting_algo}}
We suppose that the penalized log-likelihood \eqref{penalized_likelihood} depends only on the $p(1+q\tilde{d})-$vector $\betab$ and that the vector of smoothing parameters $\gammaV$ is proposed (at each iteration of the N--R algorithm) by outer iteration and is therefore fixed in what follows.

The penalized maximum log-likelihood estimator (PMLE) $\hat{\betab}$ satisfies the following score equation
\begin{equation}
\dfrac{\partial \ell(\hat{\betab},\gammaV)}{\partial \betab} = \mathbf{X}_{\text{VAM}}^\T \mathbf{u(\hat{\betab})} - \mathbf{P}(\gammaV)\hat{\betab} = \boldsymbol{0}, \notag
\end{equation}
where $\mathbf{u(\betab)}= \partial \ell(\betab)/\partial \etab \in \mathbb{R}^{p n_\mathbf{r}}$ and $\mathbf{X}_{\text{VAM}}$ is as defined in \eqref{VGAM_eta}. 
To obtain $\hat{\betab}$, we update $\betab^{(a-1)}$, the $(a-1)$th estimate of the true $\betab_0$, by Newton--Raphson:
\begin{equation}
\betab^{(a)} = \betab^{(a-1)} + \mathbf{I}\left( \betab^{(a-1)}\right) ^{-1} \left\lbrace \mathbf{X}_{\text{VAM}}^\T \mathbf{u}(\betab^{(a-1)}) - \mathbf{P}(\gammaV) \betab^{(a-1)} \right\rbrace, \label{NR_step}
\end{equation}
where 
\begin{eqnarray}
\begin{cases}\mathbf{I}\left( \betab^{(a-1)}\right)  = - \dfrac{\partial^2 \ell(\betab,\gammaV)}{\partial \betab \partial \betab^\T} = \mathbf{X}_{\text{VAM}}^\T \mathbf{W}(\betab^{(a-1)}) \mathbf{X}_{\text{VAM}} + \mathbf{P}(\gammaV),  \notag \\
\mathbf{W}(\betab^{(a-1)}) = - \dfrac{\partial^2 \ell(\betab)}{\partial \etab \partial \etab^\T} \in \mathbb{R}^{p n_\mathbf{r} \times p n_\mathbf{r}}. \notag
\end{cases}
\end{eqnarray}
The matrix $\mathbf{W}(\betab^{(a-1)})$ is termed the working weight matrix. If the expectation ${\rm{E}}\{\partial^2 \ell(\betab)/\partial \etab \partial \etab^\T \}$ is obtainable, a Fisher scoring algorithm is then preferred, as it ensures the positive definiteness of $\mathbf{W}$ over a larger region of the parameter space $\mathbf{B}$ than in the N--R algorithm. When the working weight matrix is not positive definite, which might happen when the parameter $\betab^{(a-1)}$ is far from the true $\betab_0$, a Greenstadt \citep{Greenstadt} modification is applied, and the negative eigenvalues of $\mathbf{W}(\betab^{(a-1)})$ are replaced by their absolute values. With the different families of angular densities considered in Examples~\ref{ex:logistic_surf}--\ref{ex:pairwise_beta}, the expected information matrix is not obtainable and is hence replaced by the observed information matrix on which a Greenstadt modification is applied whenever needed. See \citet[Section 9.2]{Yee_2015} for other remedies and techniques for deriving well-defined working weight matrices.

Let $\mathbf{z}^{(a-1)} := \mathbf{X}_{\text{VAM}} \betab^{(a-1)} + \mathbf{W}(\betab^{(a-1)})^{-1} \mathbf{u}(\betab^{(a-1)})$ be the $p n_\mathbf{r}-$vector of working responses. Then, \eqref{NR_step} can be rewritten in a PIRLS form as
\begin{eqnarray}
\betab^{(a)} &=&\left\lbrace \mathbf{X}_{\text{VAM}}^\T \mathbf{W}(\betab^{(a-1)}) \mathbf{X}_{\text{VAM}} + \mathbf{P}(\gammaV)\right\rbrace ^{-1} \mathbf{X}_{\text{VAM}}^{T} \mathbf{W}(\betab^{(a-1)}) \mathbf{z}^{(a-1)} \notag \\
&=& \left\lbrace \mathbf{X}_{\text{PVAM}}^\T \tilde{\mathbf{W}}^{(a-1)} \mathbf{X}_{\text{PVAM}}\right\rbrace ^{-1} \mathbf{X}_{\text{PVAM}}^\T \tilde{\mathbf{W}}^{(a-1)} \mathbf{y}^{(a-1)}, \notag
\end{eqnarray}
where $\mathbf{X}_{\text{PVAM}}$, $\mathbf{y}^{(a-1)}$, and $\tilde{\mathbf{W}}^{(a-1)}$ are augmented versions of $\mathbf{X}_{\text{VAM}}$, $\mathbf{z}^{(a-1)}$ and $\mathbf{W}(\betab^{(a-1)})$, respectively, and are defined as
\begin{eqnarray}
\begin{cases}
\mathbf{X}_{\text{PVAM}} = \begin{pmatrix}
\mathbf{X}_{\text{VAM}} & \tilde{\mathbf{X}}
\end{pmatrix}^\T \in \mathbb{R}^{p(1+ n_\mathbf{r} + q\tilde{d}) \times p(1+ q\tilde{d})}, \notag \\
\mathbf{y}^{(a-1)} = \begin{pmatrix}
\mathbf{z}^{(a-1)} & \boldsymbol{0}_{p(1+q\tilde{d})}
\end{pmatrix}^\T \in \mathbb{R}^{p(1+ n_\mathbf{r} + q\tilde{d})},  \notag \\
\tilde{\mathbf{W}}^{(a-1)} = \text{diag}\left( \mathbf{W}(\betab^{(a-1)}), \mathbf{I}_{p(1+ q\tilde{d}) \times p(1+ q\tilde{d})}\right) \in \mathbb{R}^{p(1+ n_\mathbf{r} + q\tilde{d}) \times p(1+ n_\mathbf{r} + q\tilde{d})}. \notag
\end{cases}
\end{eqnarray}
\noindent The algorithm stops when the change in the coefficients $\betab$ between two successive iterations is sufficiently small. Convergence of the N--R algorithm is not guaranteed and might not occur if the quadratic approximation of $\ell(\betab,\gammaV)$ around $\hat{\betab}$ is poor. See \cite{Yee_2015,Yee_2016} for more details.

The plug-in penalized maximum log-likelihood estimator of the covariate-dependent angular density is defined as
\begin{equation}
 \widehat{h}_{\mathbf{x}}(\mathbf{w}) \equiv h\lbrace \mathbf{w}; \mathbf{g}^{-1}(\mathbf{X}_{\text{VAM}} \hat{\betab})\rbrace . \label{estim_plug}
\end{equation}

In the following section, we give details about the selection of the smoothing parameters $\gammaV$, which is outer to the PIRLS algorithm.

\subsection{\textsf{Selection of the Smoothing Parameters}\label{subsec:selection_gamma}}
To implement the PIRLS algorithm performed at each iteration of the N--R algorithm, a smoothing parameter selection procedure is conducted by minimizing a prediction error estimate given by the generalized cross validation (GCV) score. 

Let $\mathbf{A}^{(a-1)}(\gammaV)$ be the influence matrix of the fitting problem at the $a$th iteration, defined as
\begin{equation}
\mathbf{A}^{(a-1)}(\gammaV) = \mathbf{X}_{\text{PVAM}} \left\lbrace \mathbf{X}_{\text{PVAM}}^\T \tilde{\mathbf{W}}^{(a-1)} \mathbf{X}_{\text{PVAM}}\right\rbrace ^{-1} \mathbf{X}_{\text{PVAM}}^\T \tilde{\mathbf{W}}^{(a-1)}. \notag
\end{equation}
Then, by minimizing the GCV score
\begin{equation}
\text{GCV}^{(a-1)} = \dfrac{n_{\mathbf{r}} \left\lbrace \mathbf{y}^{(a-1)} - \mathbf{A}^{(a-1)}(\gammaV) \mathbf{y}^{(a-1)}\right\rbrace ^\T \tilde{\mathbf{W}}^{(a-1)} \left\lbrace \mathbf{y}^{(a-1)} - \mathbf{A}^{(a-1)}(\gammaV) \mathbf{y}^{(a-1)}\right\rbrace}{\left[n_{\mathbf{r}} -  {\rm{trace}}\left\lbrace \mathbf{A}^{(a-1)}(\gammaV)\right\rbrace \right] ^2}, \notag
\end{equation}
we aim at balancing between goodness of fit and complexity of the model, which is measured by the trace of the influence matrix and termed the effective degrees of freedom (EDF). The EDF of the fitted VGAM \eqref{VGAM_eta} are defined as the EDF obtained at convergence, that is, ${\rm{trace}}\left\lbrace \mathbf{A}^{(c-1)}(\gammaV)\right\rbrace,$ where $c$ is the iteration at which convergence occurs.

Both the fitting algorithm of Section~\ref{subsec:fitting_algo} and the smoothing parameter selection are implemented in the \texttt{R} package \texttt{VGAM} \citep{package_VGAM}, with the latter being required from the \texttt{R} package \texttt{mgcv} \citep{wood_Book_2017}.

Model selection between different, not necessarily nested, fitted VGAMs is performed based on the Akaike information criterion (AIC), where the number of parameters of the model is replaced by its EDF to account for penalization. More details on the (conditional) AIC for models with smoothers along with a corrected version of this criterion, which takes into account the smoothing parameter uncertainty, can be found in \citet[][sec.~6.11]{wood_Book_2017}.

\subsection{\textsf{Large Sample Properties}\label{subsec:large_sample_properties}}
In this section we derive the consistency and asymptotic normality of the PMLE $\hat{\betab}$ defined in Section~\ref{subsec:fitting_algo}. \\
Based on the penalized log-likelihood \eqref{penalized_likelihood}, $\hat{\betab}$ satisfies the following score equation
\begin{equation}
\mathbf{m}(\betab) - \mathbf{P}(\gammaV)\betab = \boldsymbol{0}_{p(1+q\tilde{d})}, \label{score_penalized}
\end{equation}
where $\mathbf{m}(\betab)=\partial \ell(\betab)/\partial \betab$.

Let $\mathbf{B}_0$ be an open neighborhood around the true parameter $\betab_0$. Moreover, we define $\mathbf{m}(\mathbf{y},\betab)=\partial \ell(\mathbf{y};\betab)/\partial \betab$. \\
Our asymptotic results hold under the following customary assumptions:
\begin{enumerate}[label=(\subscript{\text{A}}{\arabic*}), itemindent = 1cm]
\item $\gammaV=\begin{pmatrix}
\gamma_{(1)1} & \cdots & \gamma_{(p)1} & \cdots & \gamma_{(1)q} & \cdots & \gamma_{(p)q}
\end{pmatrix}^\T = o(n_{\mathbf{r}}^{-1/2}) \mathbf{1}_{pq}$. 
\item Regularity conditions: 
\begin{itemize}
\item If $\betab \neq \tilde{\betab}$, then $\ell(\mathbf{y};\betab) \neq \ell(\mathbf{y};\tilde{\betab})$, with $\betab, \tilde{\betab} \in \mathbf{B}$. Moreover, $\rm{E}\lbrace {\rm{sup}}_{\betab \in \mathbf{B}} \vert \ell(\mathbf{Y};\betab) \vert \rbrace < \infty$. 
\item The true parameter $\betab_0$ is in the interior of $\mathbf{B}$.
\item For $\mathbf{y} \in (0,\infty)^d$, $\ell(\mathbf{y};\betab) \in C^{3}(\mathbf{B}_0)$.
\item $\int {\rm{sup}}_{\betab \in \mathbf{B}_0} \| \mathbf{m}(\mathbf{y},\betab)\| \, \dif \mathbf{y} < \infty$ and $\int {\rm{sup}}_{\betab \in \mathbf{B}_0} \| \partial \mathbf{m} (\mathbf{y},\betab)/ \partial \betab^\T \| \dif \mathbf{y} < \infty$.
\item For $\betab \in \mathbf{B}_0$, $\mathbf{i}(\betab):={\rm{cov}} \lbrace \mathbf{m}( \mathbf{Y},\betab)\rbrace = \mathbf{X}_{\text{VAM}}^\T \mathbf{W}(\betab)\mathbf{X}_{\text{VAM}}$ exists and is positive-definite.
\item For each triplet $1 \leq q, r, s \leq p(1+q\tilde{d})$, there exists a function $M_{qrs}: (0,\infty)^d \rightarrow \mathbb{R}$ such that, for $\mathbf{y} \in (0,\infty)^d$ and $\betab \in \mathbf{B}_0$, $\vert \partial^3 \ell(\mathbf{y};\betab)/\partial \betab_{qrs} \vert \leq M_{qrs}(\mathbf{y})$, and ${\rm{E}}\left\lbrace M_{qrs}(\mathbf{Y})\right\rbrace < \infty$.
\end{itemize}
\end{enumerate}
The next theorem characterizes the large sample behavior of our estimator.
\begin{theorem}
Under $A_1$ and $A_2$, it follows that as $n_{\mathbf{r}} \to \infty$:
  \begin{enumerate}
  \item $\| \hat{\betab} - \betab_0 \| = \ensuremath{O_{p}(n_{\mathbf{r}}^{-1/2})}$. 
  \item $\sqrt{n_{\mathbf{r}}} (\hat{\betab} - \betab_0) \dto N(\mathbf{0}, \mathbf{i}(\betab_0)^{-1})$.
\end{enumerate}
\label{Theorem1}
\end{theorem}
These results are derived from a second-order Taylor expansion of the score equation \eqref{score_penalized} around the true parameter $\betab_0$ along the same lines as in \cite{Vatter_2015} and \citet[p.~147]{Davison_Stat_Models}. Similar results on the large sample behavior of the corresponding plug-in estimator \eqref{estim_plug} can be derived using the multivariate delta method. These results are useful to derive and construct approximate confidence intervals for conditional angular densities and to compare nested models based on likelihood ratio tests. Our proviso is similar to that of \cite{decarvalho2014a} in the sense that asymptotic properties of the estimator $\hat{\betab}$ are derived under the assumption of known margins and we sample from the limiting object $h_{\textbf{x}}$, whereas in practice only a sample of (estimated) pseudo-angles, $\{\widehat{\mathbf{W}}_i\}_{i = 1}^{n_{\mathbf{r}}}$, would be available. Asymptotic properties under misspecification of the parametric model set for $h_{\textbf{x}}$ could in principle be derived under additional assumptions on $\betab$ and $\mathbf{m}$, along the same lines as in standard likelihood theory \citep{knight2000}. The resulting theory is outside the scope of this work and is deliberately not studied here.

\section{\textsf{Simulation Study}}
\label{subsec:simulation}

\subsection{\textsf{Data Generating Processes and Preliminary Experiments}}
We assess the performance of our methods using the bivariate extremal dependence structures presented in Section~\ref{sec:gamspectral}---and displayed in Figure~\ref{fig:spectral_surfaces}---as well as the trivariate pairwise beta dependence model from Example~\ref{ex:pairwise_beta}---depicted in Figure~\ref{fig:pBeta_angular_density}. Monte Carlo evidence will be reported in Section~\ref{montecarlo} and in the Supplementary Materials. For now, we concentrate on illustrating the methods over a single-run experiment on these scenarios. For each dependence model from Examples~\ref{ex:logistic_surf}--\ref{ex:Husler_surf}, we draw a sample $\lbrace (Y^{i}_{1},Y^{i}_{2})\rbrace_{i=1}^n$ from the corresponding bivariate extreme value distribution $G_{x}$ with sample size $n=6000$ and where each observation $(Y^{i}_{1},Y^{i}_{2})$ has unit Fr\'{e}chet margins and is drawn from the chosen dependence model conditional on a fixed value $x^i$ of the covariate $x$. For estimating $h_x$, we only consider the observations with a radial component exceeding its 95\% quantile, and we end up with $n_{\mathbf{r}}=300$ extreme (angular) observations. To gain insight into the bias and variance of our covariate-adjusted spectral density estimator, we compute its 95\% asymptotic confidence bands based on Theorem~\ref{Theorem1} and at different values of $w \in (0,1)$. There are two possible sources of bias in our estimation procedure. First, the limiting extremal dependence structure is estimated at a sub-asymptotic level, i.e., based on angular observations exceeding a finite diagonal threshold level. Then, the penalization of the model likelihood causes a smoothing bias \citep{wood_Book_2017} if the smoothing parameters do not vanish at a certain rate (see Section~\ref{subsec:large_sample_properties}). The uncertainty due to the choice of the parametric model is deliberately not taken into account, that is, the simulations are performed in a well-specified framework.

\begin{figure}
  \centering
  \textbf{ \ \ \ \ \ Logistic}
  \subfloat{\includegraphics[width=0.33\textwidth]{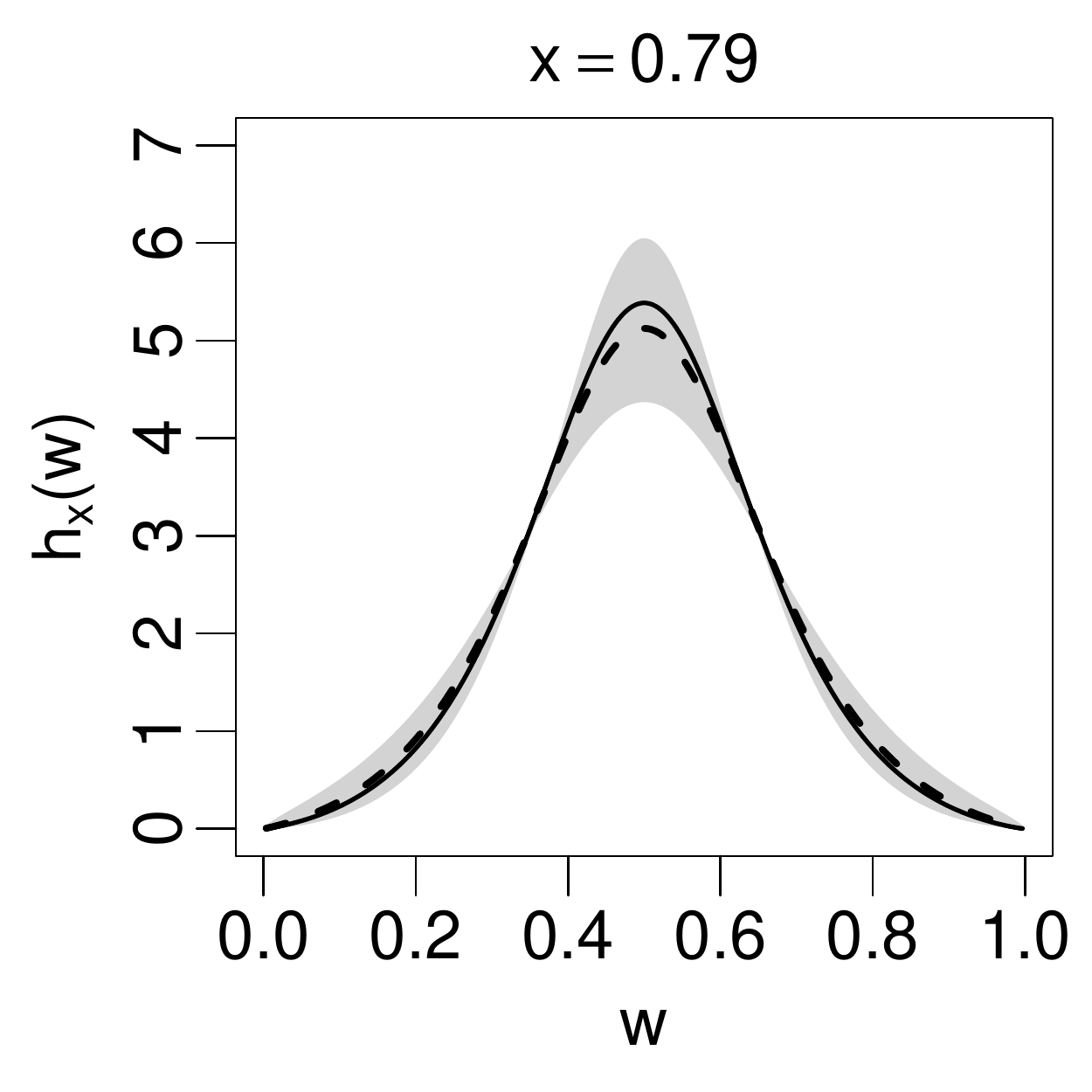}}
  \subfloat{\includegraphics[width=0.33\textwidth]{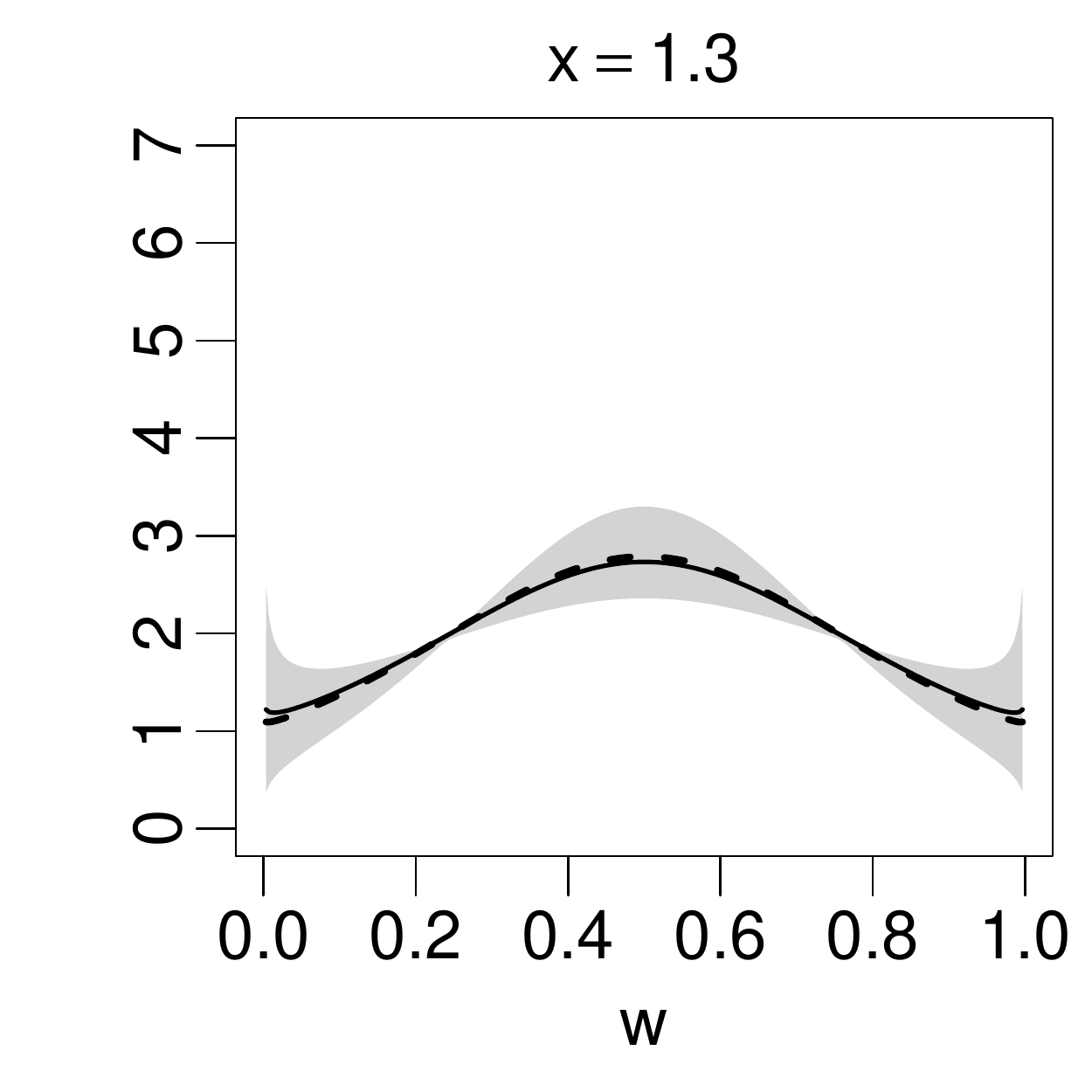}}
  \subfloat{\includegraphics[width=0.33\textwidth]{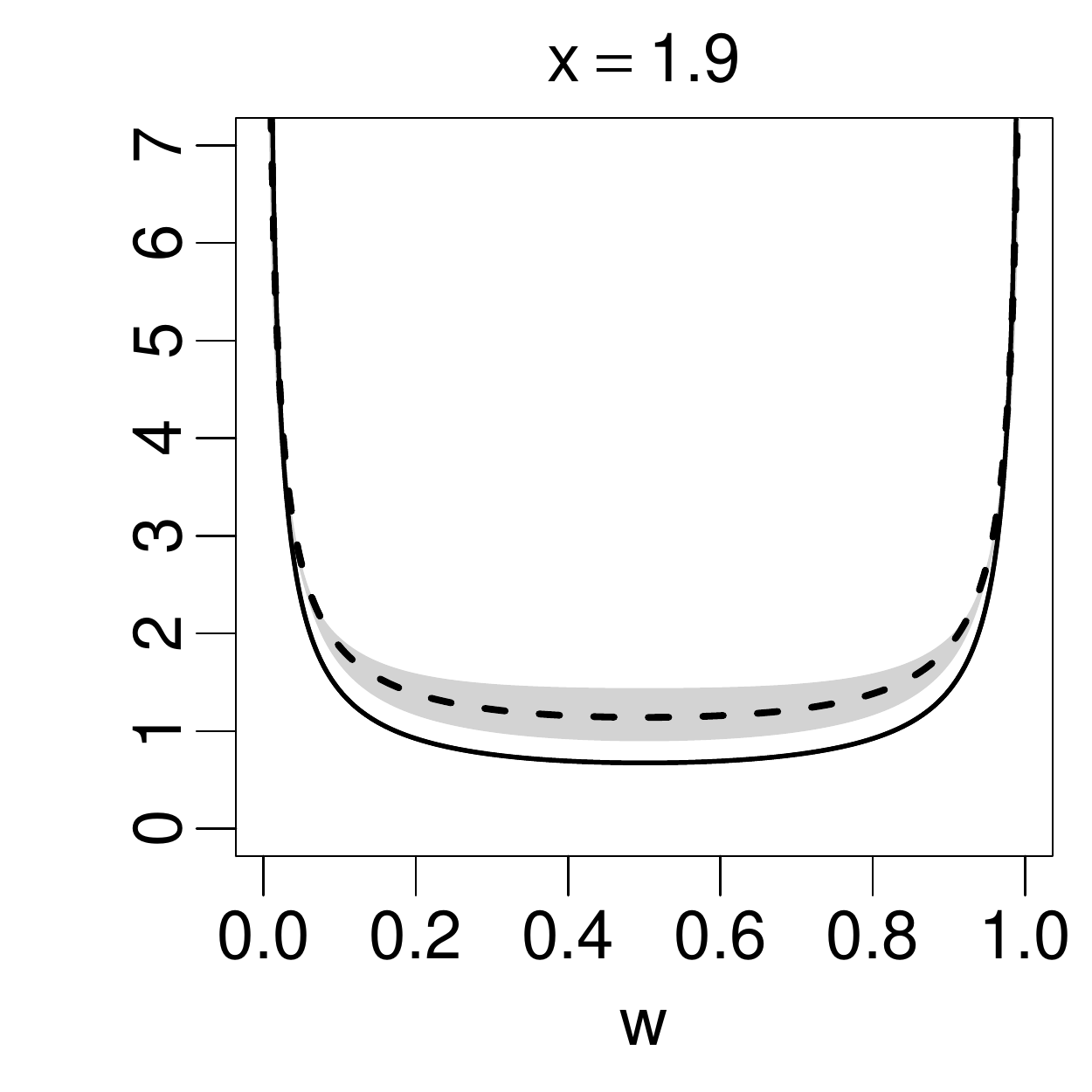}}\\
  \textbf{ \ \ \ \ \ Dirichlet}
  \hfill
  \subfloat{\includegraphics[width=0.33\textwidth]{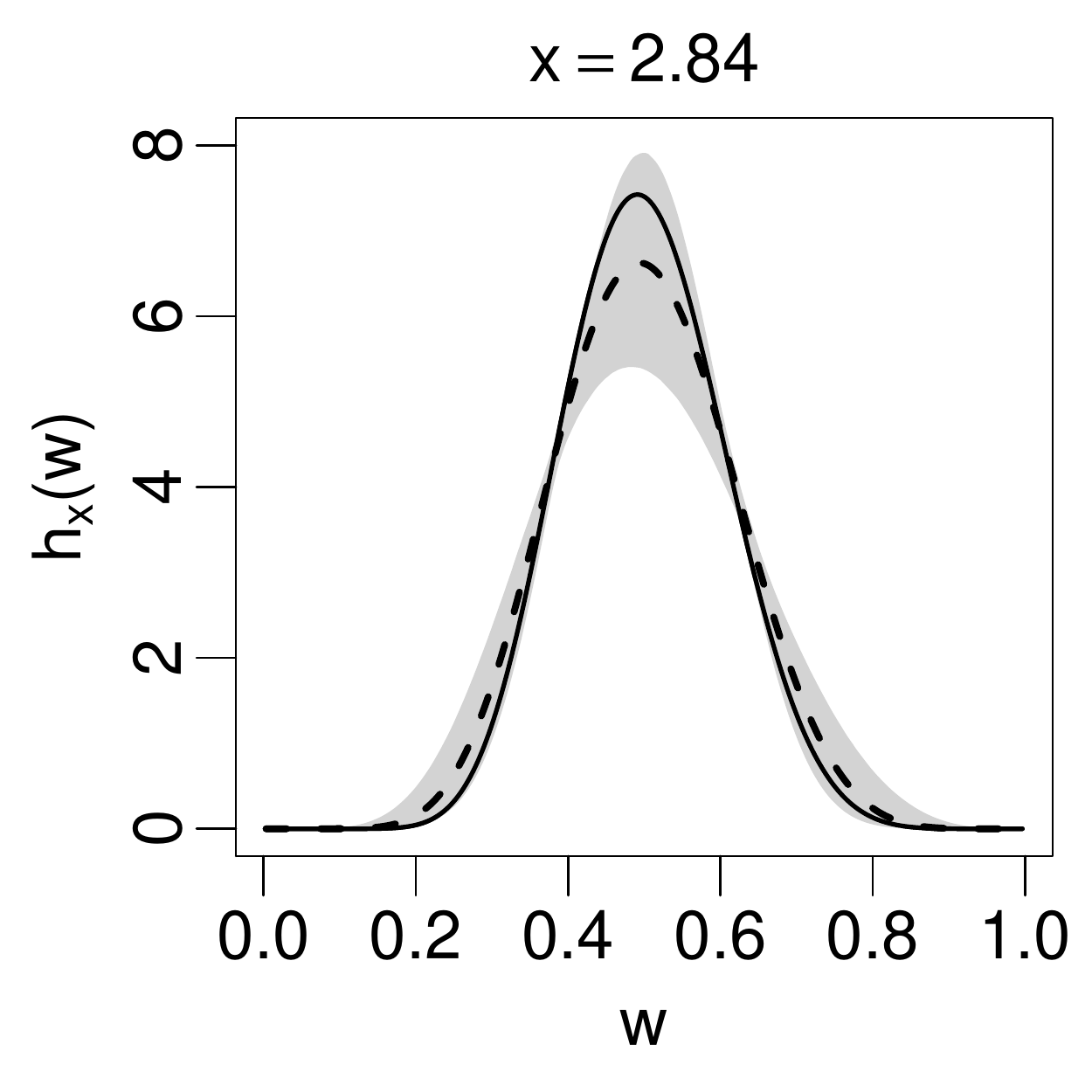}}
  \subfloat{\includegraphics[width=0.33\textwidth]{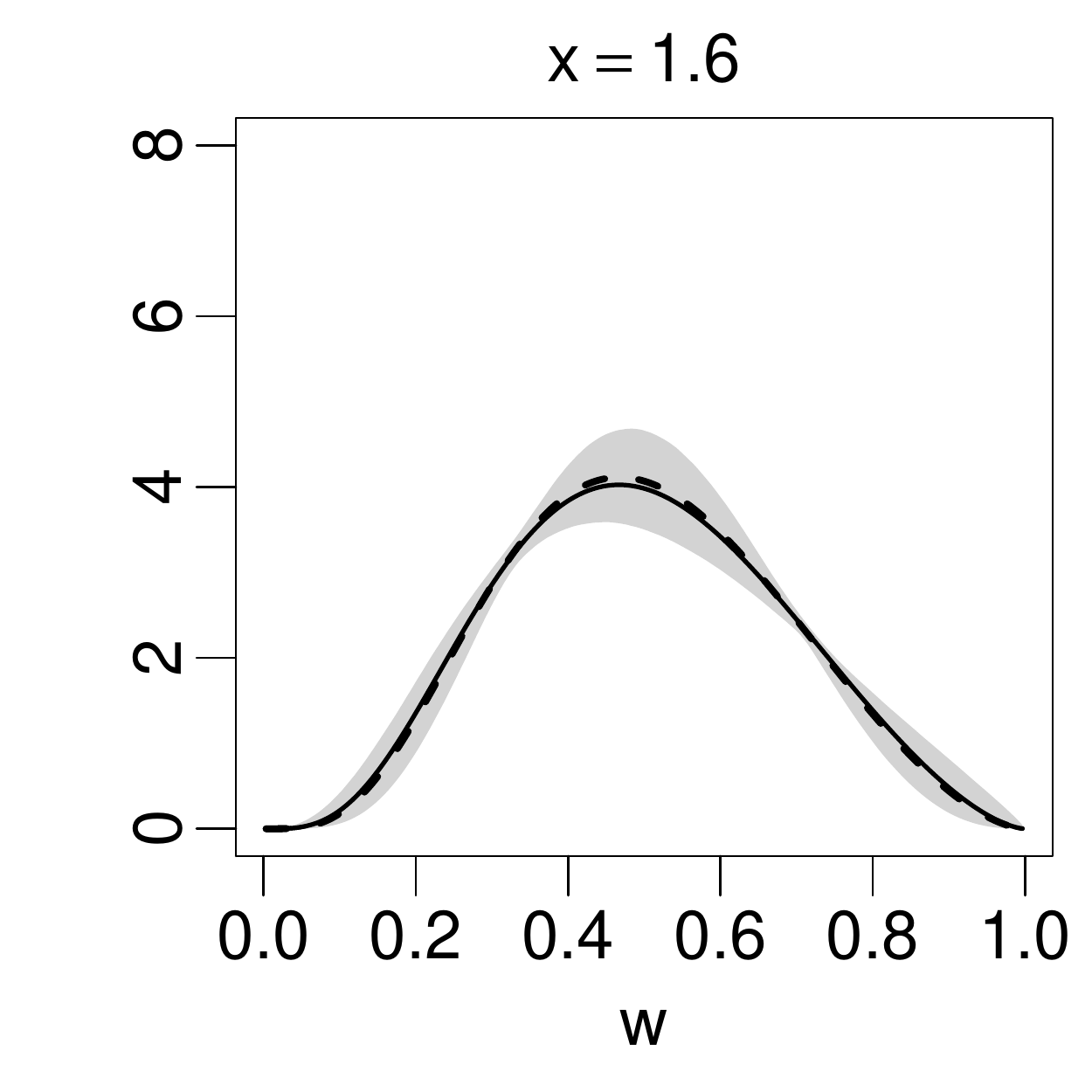}}
  \subfloat{\includegraphics[width=0.33\textwidth]{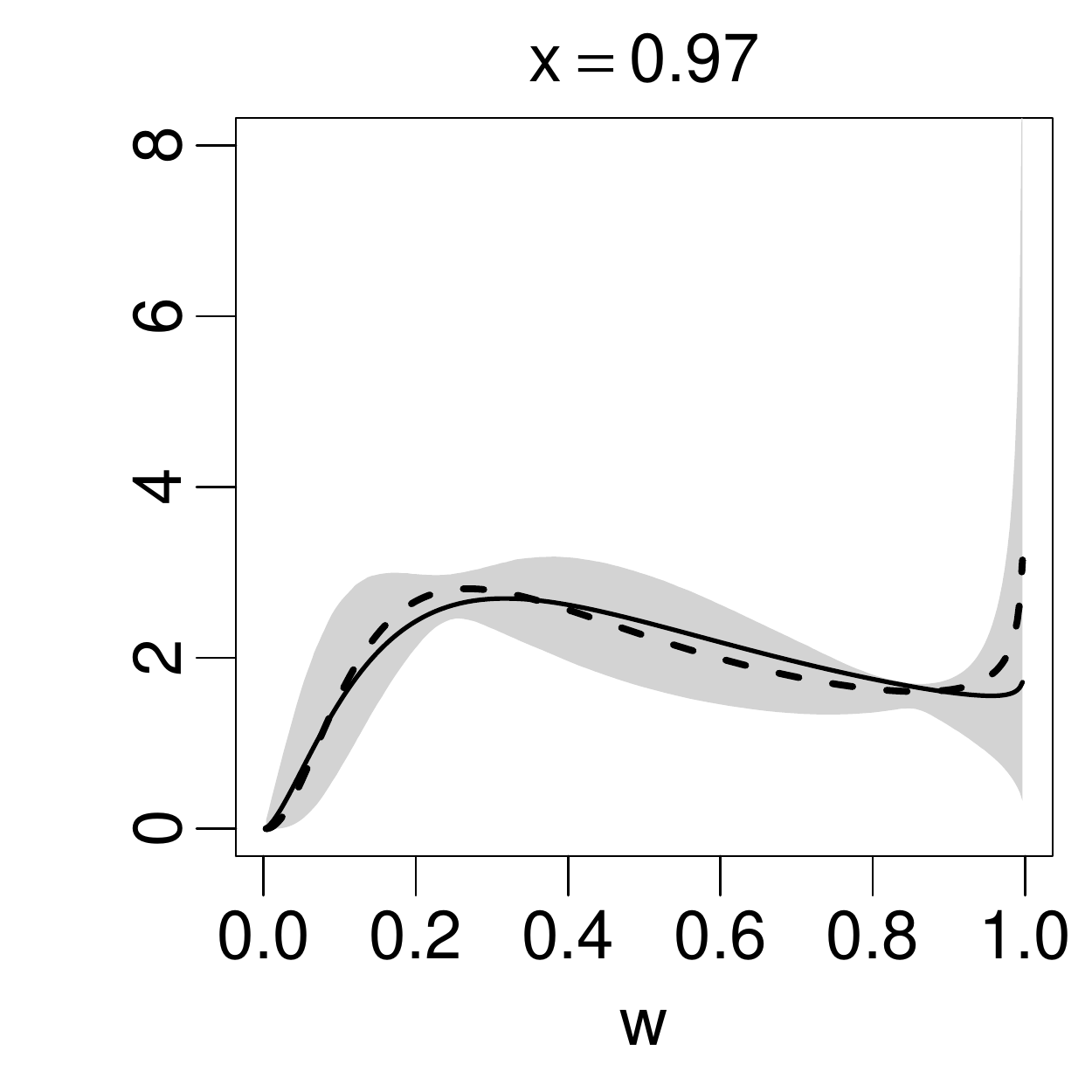}}\\
  \textbf{ \ \ \ \ H\"{u}sler--Reiss}
  \hfill
  \subfloat{\includegraphics[width=0.33\textwidth]{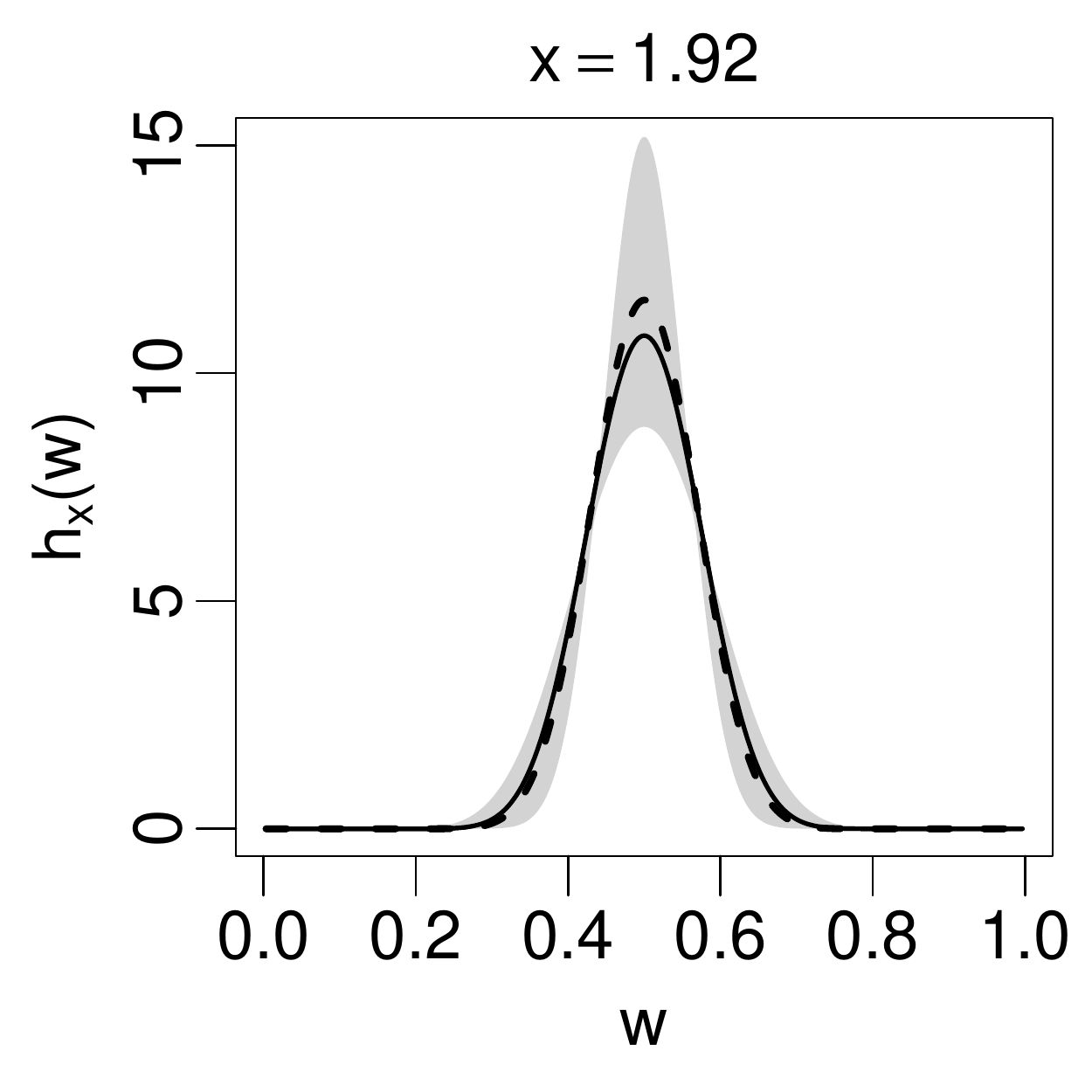}}
  \subfloat{\includegraphics[width=0.33\textwidth]{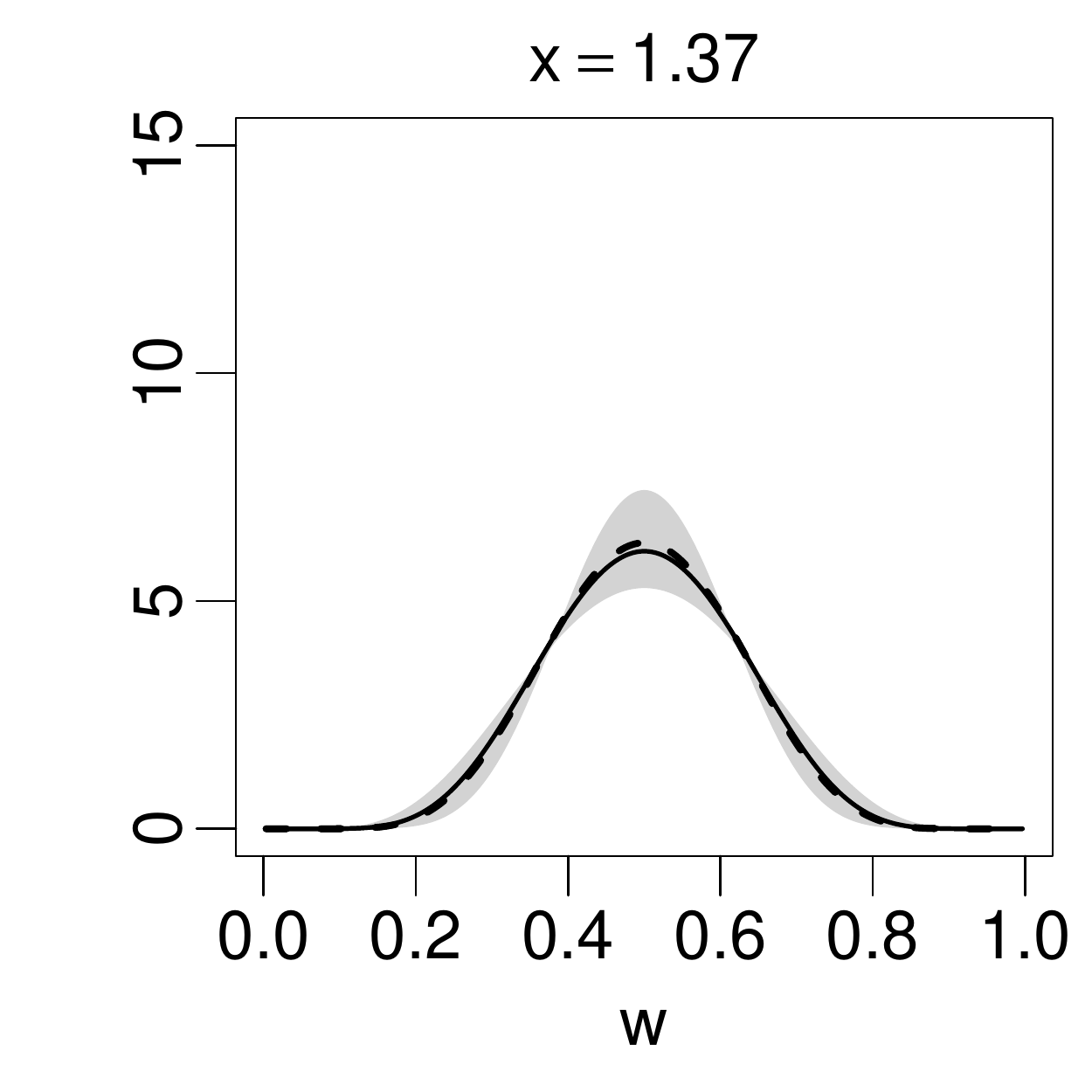}}
  \subfloat{\includegraphics[width=0.33\textwidth]{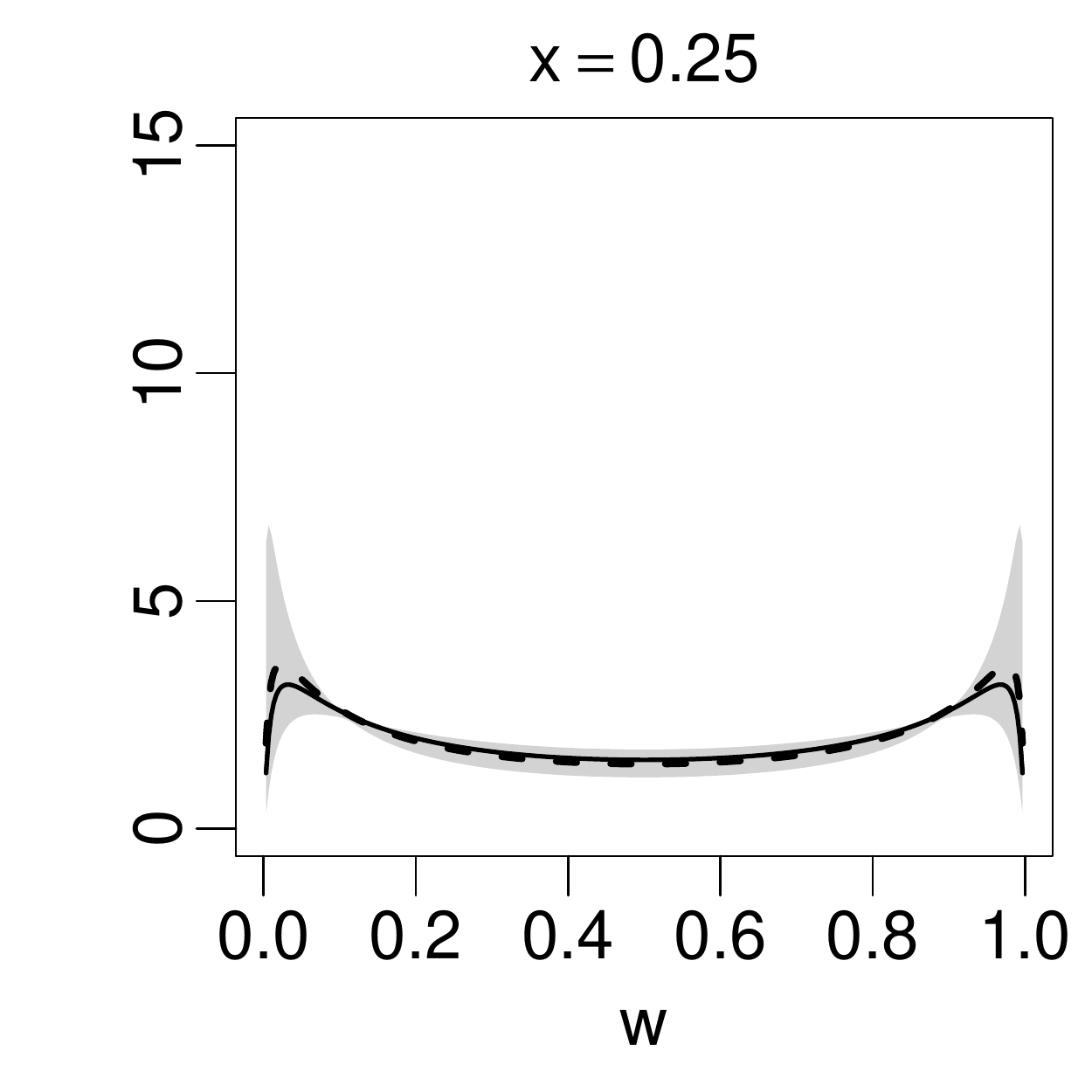}}
  \caption{\footnotesize Estimates of the covariate-adjusted spectral densities in Examples~\ref{ex:logistic_surf}, \ref{ex:Dirichlet_surf}, and \ref{ex:Husler_surf} conditional on different values of the covariate $x$ (dashed lines) along with their 95\% (pointwise) asymptotic confidence bands (grey area). The true spectral densities are displayed in solid lines.\label{simu_bilogistic}}
\end{figure}

Figure~\ref{simu_bilogistic} displays the estimates of the covariate-adjusted spectral densities from Examples~\ref{ex:logistic_surf}, \ref{ex:Dirichlet_surf}, and \ref{ex:Husler_surf} for various fixed values of the covariate $x$ that induce different extremal dependence strengths.
All panels show that for the different extremal dependence schemes (strength and asymmetry), the covariate-adjusted spectral densities are accurately estimated and the true curves fall well within the 95\% confidence bands. A systematic slight upward bias is observed when approaching extremal independence. This is due to the residual dependence in the data that we observe at finite threshold levels but that should vanish at an asymptotic level. This issue can be corrected either by taking higher threshold levels or considering angular observations simulated from the true spectral density. Finally, the estimates in the Dirichlet case seem to be a bit more biased, and this might be explained by the fact that both of the two non-orthogonal parameters of the model depend smoothly on the covariate $x$.

\begin{figure}
  \centering
  \subfloat{\includegraphics[width=0.3\textwidth]{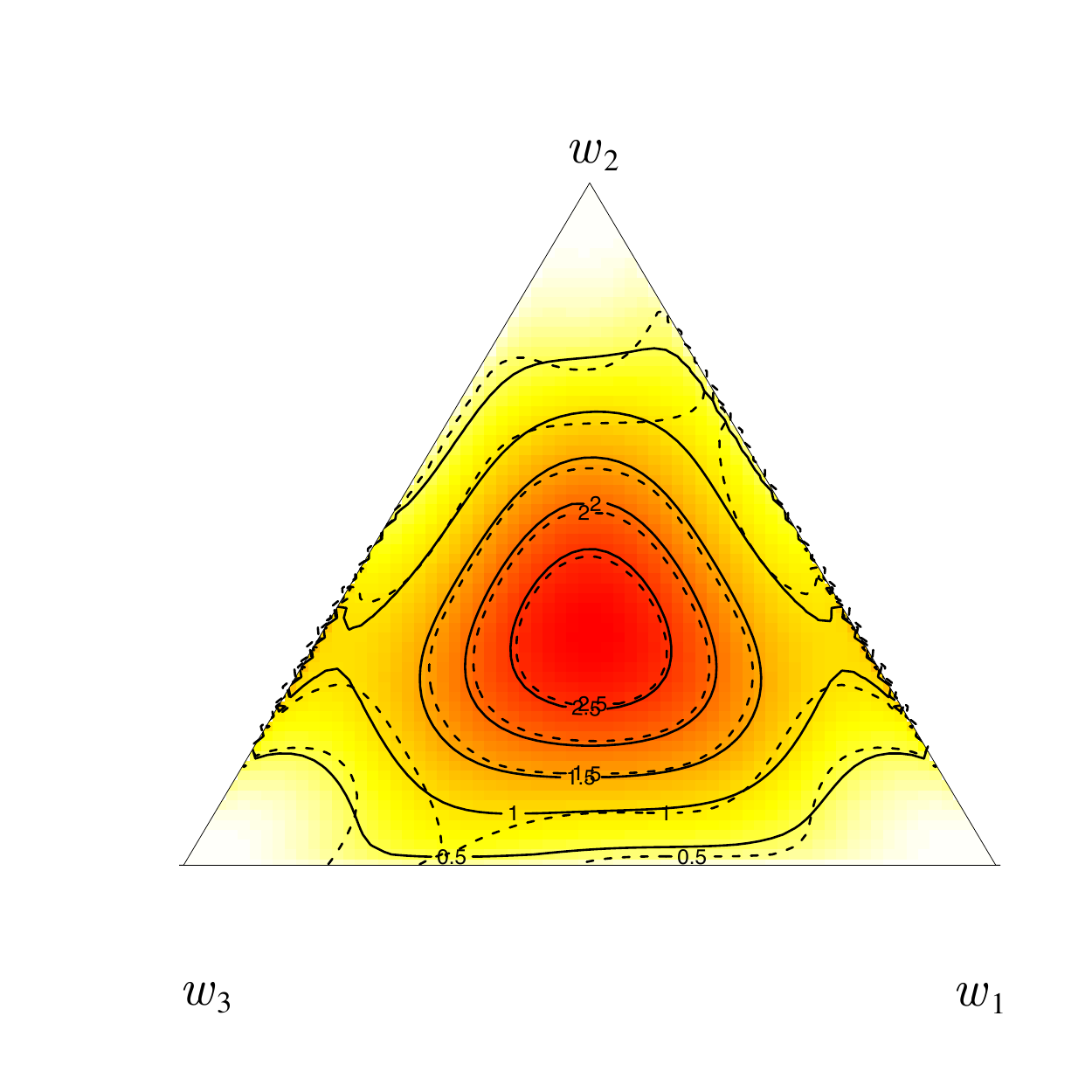}}
  \subfloat{\includegraphics[width=0.3\textwidth]{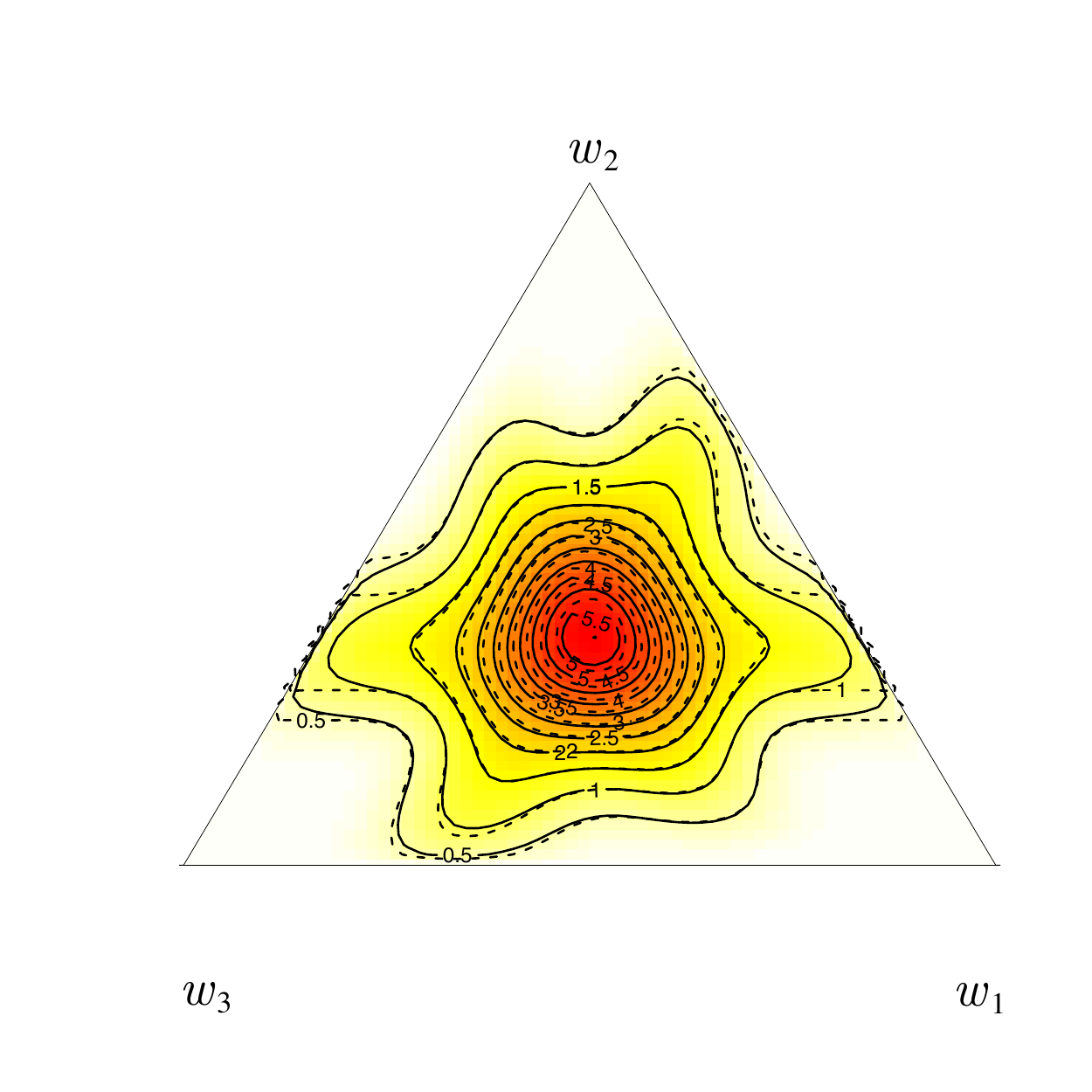}}
  \subfloat{\includegraphics[width=0.3\textwidth]{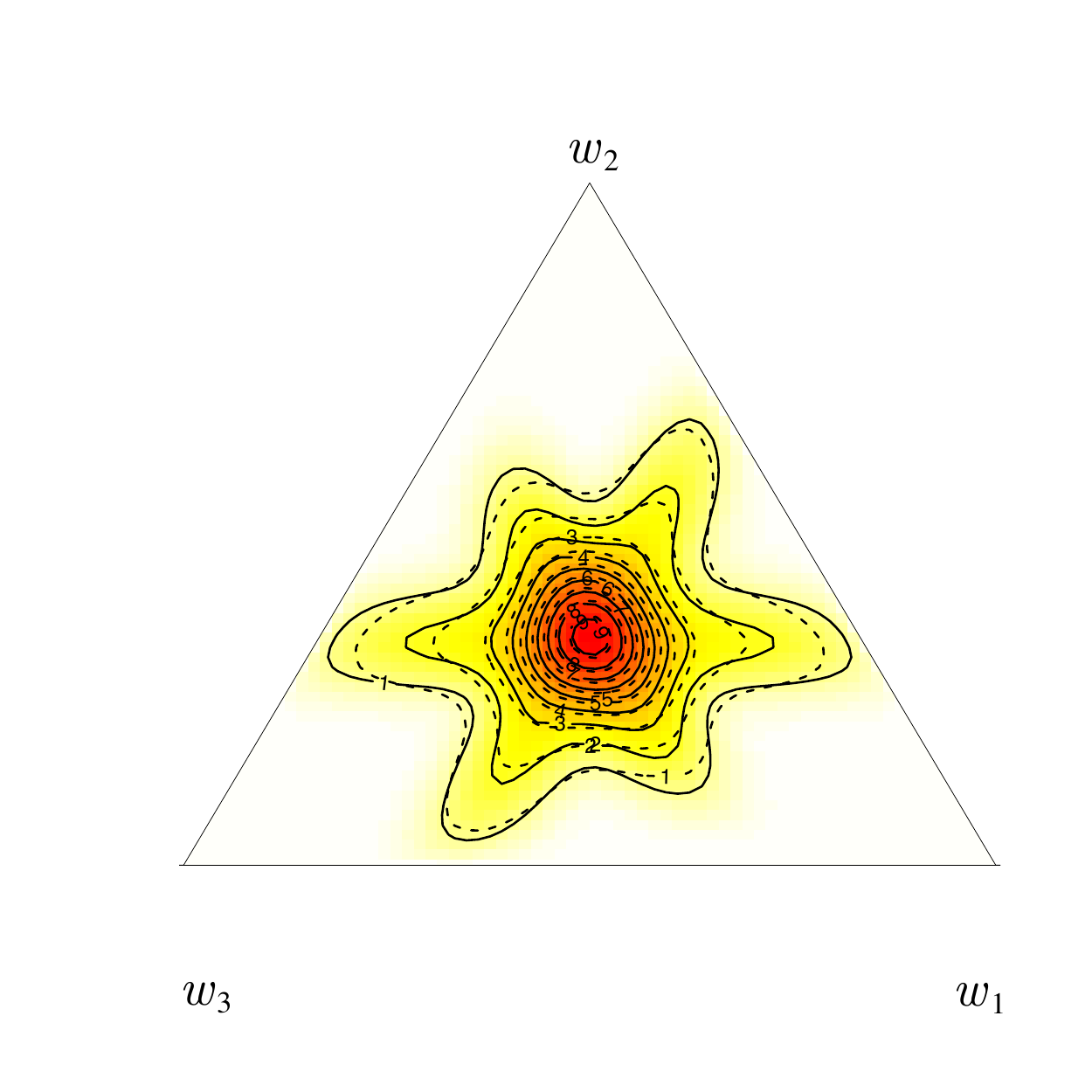}}
  \caption{\footnotesize Contour plots of the covariate-adjusted pairwise beta spectral density estimate (dashed lines) at $x=1.5$ (left), $x=2.46$ (middle), and $x=3.22$ (right). The contour plots of the true spectral density are displayed in solid lines. \label{pairwise_beta_VGAM_contours}}
\end{figure}

We now consider the case of the trivariate pairwise beta dependence model from Example~\ref{ex:pairwise_beta}. The construction of the pairwise beta covariate-adjusted spectral density---which extends \cite{pairwise.beta.Cooley}---is such that the corresponding multivariate extreme value distribution cannot be computed in closed form. Hence, we draw a sample $\left\lbrace \left( w_{i,1},w_{i,2},w_{i,3} \right)\right\rbrace _{i=1}^{n_{\mathbf{r}}}$ with sample size $n_{\mathbf{r}}=300$ where each observation $\left( w_{i,1},w_{i,2},w_{i,3} \right)$ is drawn from the pairwise beta model conditional on a fixed value $x_i$ of the covariate $x$, as illustrated in Figure~\ref{fig:pBeta_angular_density}. Figure~\ref{pairwise_beta_VGAM_contours} displays the contour plots of the estimates of the covariate-adjusted spectral density from Example~\ref{ex:pairwise_beta} at three fixed values of $x$.
All panels in Figure~\ref{pairwise_beta_VGAM_contours} show that, for the different extremal dependence schemes, i.e., for the different considered values of $x$, the contour plots of the estimates are remarkably close to the actual contour plots. The estimates are slightly more biased near the edges of the simplex than in the center, reflecting a better estimation of the global dependence parameter compared to the pairwise dependence parameters. 

\subsection{\textsf{Monte Carlo Evidence}}\label{montecarlo}
A Monte Carlo study was conducted by simulating $500$ samples of sizes $6000$ and $10000$, that is, $n_{\mathbf{r}}=300$ and $n_{\mathbf{r}}=500$ extreme (angular) observations, respectively. As can be seen from Figures 1 and 2 in the Supplementary Materials, our method successfully recovers the corresponding target covariate-adjusted angular densities with a high level of precision over the simulation study. In what follows we focus on documenting how the level of accuracy increases when the number of observations increases by assessing the mean integrated absolute error (MIAE)---which for the bivariate case can be written as
\begin{equation}\label{miae}
  \text{MIAE} = \text{E} \left\lbrace \int_{\mathcal{X}} \int_{0}^{1} |\widehat{h}_x(w) - h_x(w)| \, \dif w \, \dif x \right\rbrace. \notag
\end{equation}
The results are reported in Table~\ref{MIAE}.
\begin{table}[H]
\caption{\footnotesize Mean integrated absolute error (MIAE) estimates computed from $500$ samples for the covariate-adjusted spectral densities in Examples~\ref{ex:logistic_surf}--\ref{ex:Husler_surf}; $n_{\mathbf{r}}$ denotes the number of angular observations.}
  \centering
\begin{tabular}{lcc} 
 \hline  \hline 
$n_{\mathbf{r}}$ & Covariate-adjusted angular density & MIAE \\
\hline
\multirow{3}{5em}{$300$} & Logistic & $0.3936$ \\ 
& Dirichlet & $0.3337$ \\ 
& H\"{u}sler--Reiss & $0.2463$ \\ 
\hline
\multirow{3}{5em}{$500$} & Logistic & $0.3538$ \\ 
& Dirichlet & $0.2600$ \\ 
& H\"{u}sler--Reiss & $0.2016$ \\ 
\hline
\end{tabular}
  \label{MIAE}
\end{table}
\noindent As expected, an increase in the number of angular observations leads to a reduction of MIAE. Evidence from Table~\ref{MIAE} should be supplemented with Figures 1 and 2 in the Supplementary Materials. The latter offer a more granular level of detail than that of Table~\ref{MIAE} on the behavior of the estimator over specific values of the covariate and of the unit simplex.

\section{\textsf{Extreme Temperature Analysis}}
\label{sec:application}
\subsection{\textsf{Data Description, Motivation for the Analysis, and Preprocessing} \label{data_description}}
In this section, we describe an application to modeling the dependence between extreme air winter (December--January--February) temperatures at two sites in the Swiss Alps: Montana---at an elevation of $1427$m---and Zermatt---at an elevation of $1638$m. The sites are approximatively $37$km apart. \\
In the Alpine regions of Switzerland, there is an obvious motivation to focus on extreme climatic events, as their impact on the local population and infrastructure can be very costly. As stated by \cite{Beniston_2007}, warm winter spells, that is, periods with strong positive temperature exceedances in winter, can exert significant impacts on the natural ecosystems, agriculture, and water supply:
\begin{quote}\footnotesize
 ``Temperatures persistently above $0^{\circ}$C will result in early snow-melt and a shorter seasonal snow cover, early water runoff into river basins, an early start of the vegetation cycle, reduced income for alpine ski resorts and changes in hydro-power supply because of seasonal shifts in the filling of dams \citep{Beniston_2004}.''
\end{quote}
In this analysis, we are interested in the dynamics of the dependence between extreme air temperatures in Montana and Zermatt during the winter season. The dynamics of both extreme high and extreme low winter temperatures in these two sites will be assessed and linked to the following explanatory factors: time (in years) ($t$), day within season ($d$), and the NAO (North Atlantic Oscillation) index ($z$); the latter is a normalized pressure difference between Iceland and the Azores that is known to have a major direct influence on the alpine region temperatures, especially during winter \citep{Beniston_2005}. The choice of the studied sites is of great importance in this analysis. \cite{Beniston1996} showed that both cold and warm winters exhibit temperature anomalies that are altitude-dependent, with high-elevation resorts being more representative of free atmospheric conditions and less likely to be contaminated by urban effects. Therefore, to study the ``pure" effect of the above-mentioned explanatory covariates on the winter temperature extremal dependence, we choose the two high elevation sites Montana and Zermatt.

The data consist of daily winter temperature minima and maxima measured at $2$m above ground surface and were obtained from the MeteoSwiss website (\url{www.meteoswiss.admin.ch}). The data were available from 1981 to 2016, giving a total of $3190$ winter observations per site. Daily NAO index measurements were obtained from the NOAA (National Centers for Environmental Information), at \url{https://www.ngdc.noaa.gov/ftp.html}.

\begin{figure}[!h]
  \centering
  \subfloat{\includegraphics[width=0.3\textwidth]{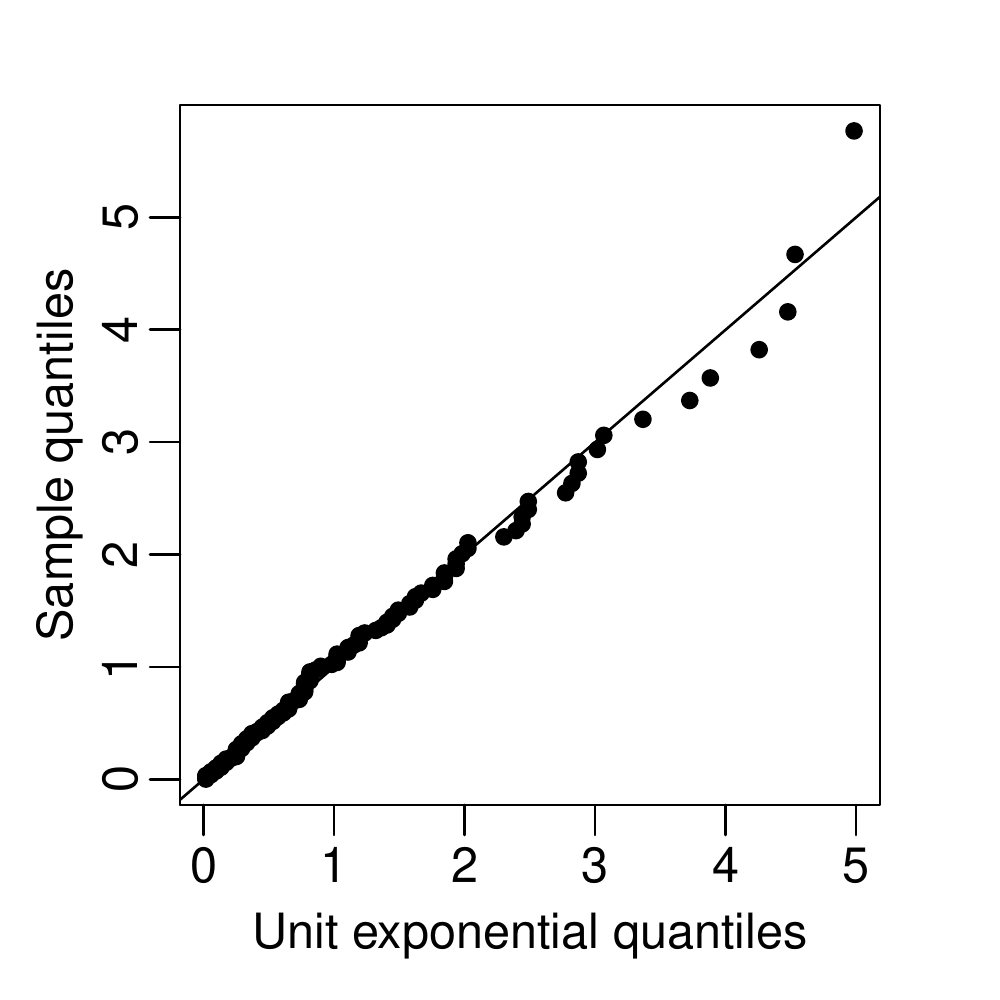}}
  \subfloat{\includegraphics[width=0.3\textwidth]{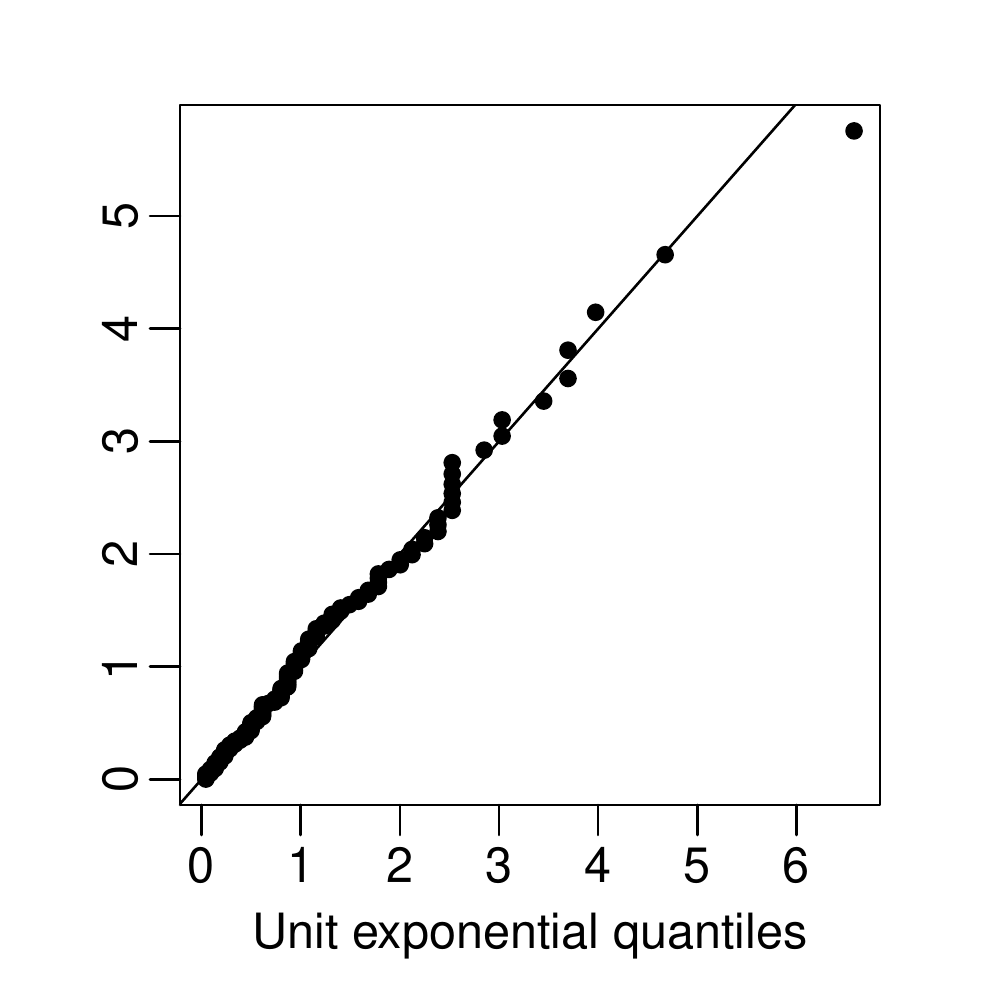}} \\
  \subfloat{\includegraphics[width=0.3\textwidth]{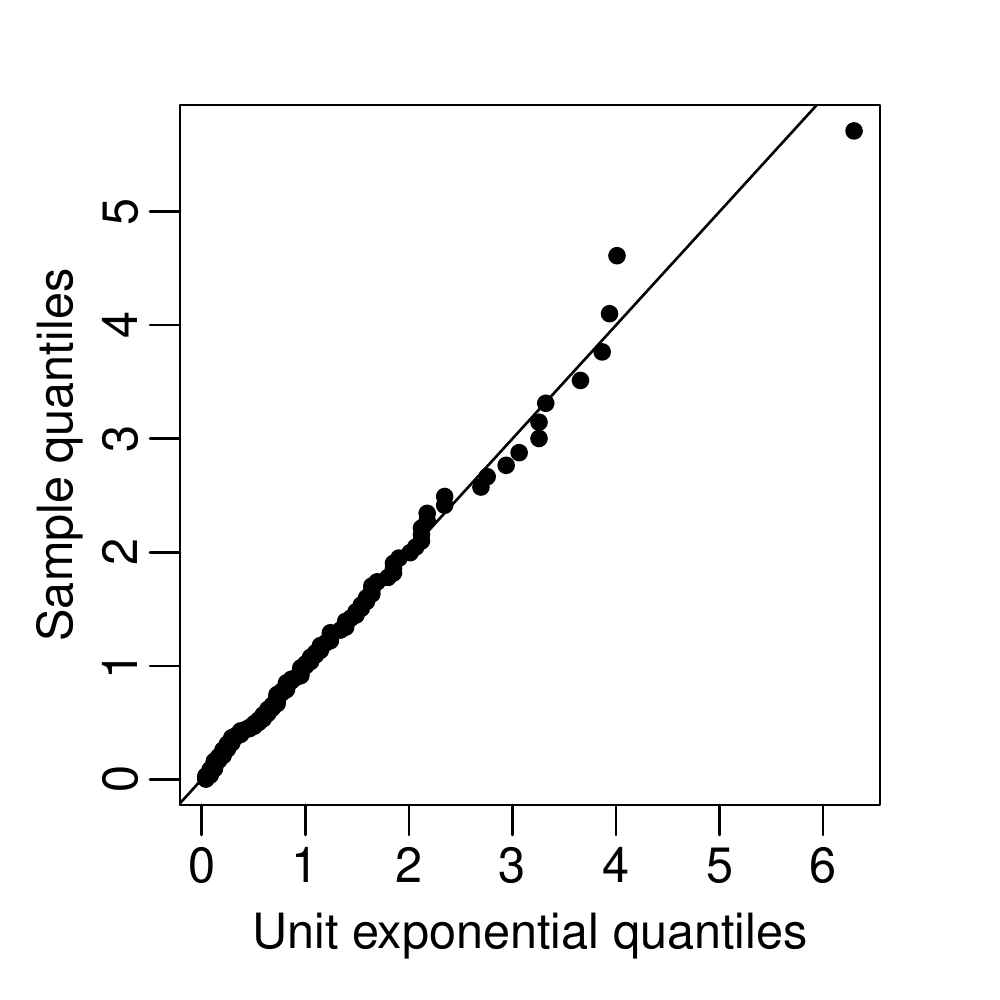}}
  \subfloat{\includegraphics[width=0.3\textwidth]{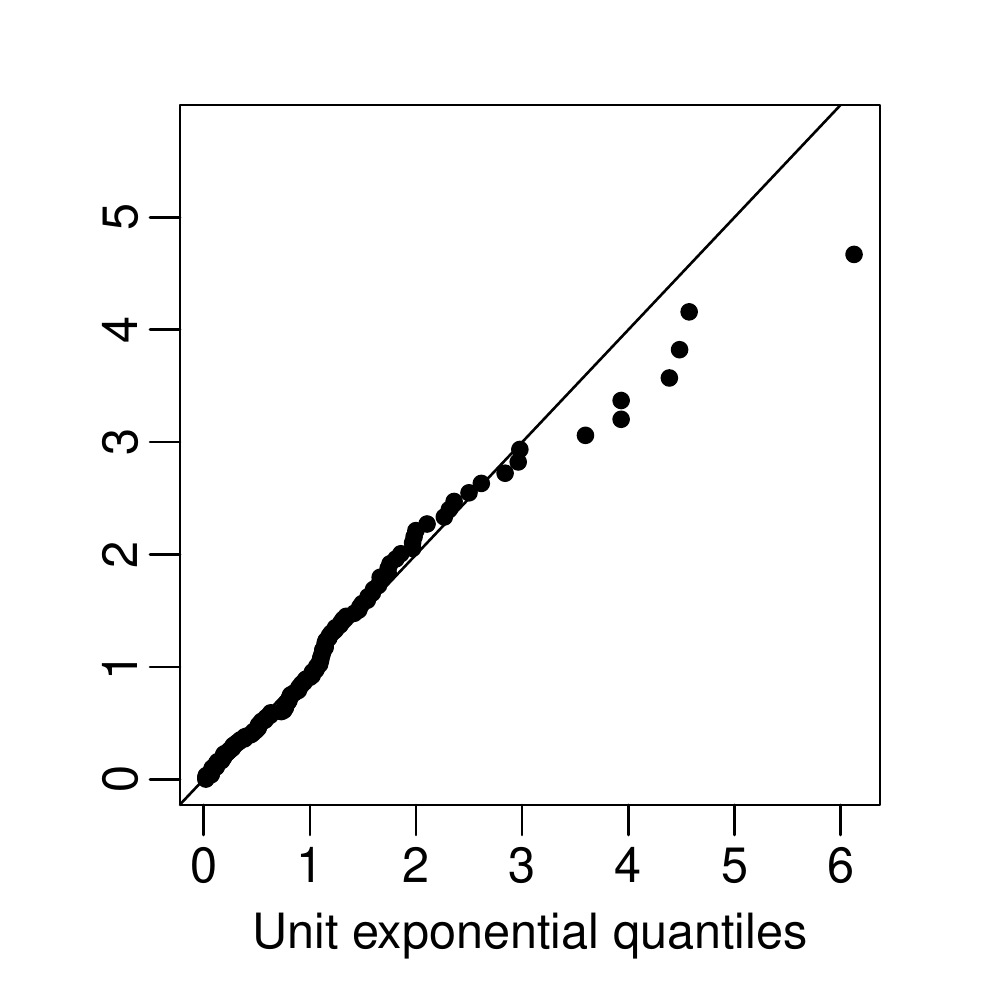}}
  \caption{\footnotesize Diagnostic plots of the GPD modeling of the threshold exceedances of the daily maximum winter temperatures (left) and the daily minimum winter temperatures (right) in Montana (top) and Zermatt (bottom).\label{qq_plots_GPD}}
\end{figure}

We first transform the minimum temperature data by multiplication by $-1$ and then fit at each site---and to both daily minimum and maximum temperatures---a Generalized Pareto Distribution (GPD) \citep[ch.~4]{coles2001}
\begin{equation}
\text{G}_{\sigma,\xi}(y) = 1- \left( 1 + \xi \dfrac{y}{\sigma}\right) _{+}^{-1/\xi} \label{GPD},
\end{equation}
to model events above the $95\%$ quantile $u_{95}$ for each of the four temperature time series. In \eqref{GPD}, $\sigma>0$ is the scale parameter that depends on $u_{95}$, and $-\infty < \xi < \infty$ is the shape parameter. As is common with temperature data analysis, we test the effect of time $t$ on the behavior of the threshold exceedances by allowing the scale parameter of the GPD \eqref{GPD} to smoothly vary with $t$ \citep{chavez_davison}. Based on the likelihood ratio tests, a model with a non-stationary scale parameter is preferred only in Zermatt for the threshold exceedances of the daily minimum temperatures ($p$-value $\approx 0.022$). Graphical goodness-of-fit tests for the four GPD models are conducted by comparing the distribution of a test statistic $S$ with the unit exponential distribution (if $Y \sim G_{\sigma,\xi}$, then $S = - \ln\lbrace 1-G_{\sigma,\xi}(Y)\rbrace$ is unit exponentially distributed). Figure~\ref{qq_plots_GPD} displays the resulting qq-plots and confirms the validity of these models. 
The fitted models are then used to transform the data to a common unit Fr\'{e}chet scale by probability integral transform and where the empirical distribution is used below $u_{95}$. This results in two datasets of bivariate observations (in Montana and Zermatt) with unit Fr\'{e}chet margins: one for the daily maximum temperatures and the other one for the daily minimum temperatures.

Following the theory developed in Section~\ref{sec:preps}, we transform each of the two datasets into pseudo-datasets of radial and angular components. By retaining the angular observations corresponding to a radial component exceeding its $95\%$ quantile in each pseudo-dataset, we end up with two pseudo-samples of $160$ extreme bivariate (angular) observations in each pseudo-dataset.

\subsection{\textsf{Covariate-Adjusted Dependence of Extreme Temperatures} \label{extremal_dep_application}}
In the following analyses of the dynamics of the dependence between extreme temperatures in Montana and Zermatt---and in line with findings from previous analyses of extreme temperatures in Switzerland \citep{Davison2011,Davison_Huser_Thibaut,Dombry_2013}---we assume asymptotic dependence in both extremely high and extremely low winter temperatures. 
\subsubsection*{Dependence of Extreme High Winter Temperatures}
The covariate-adjusted bivariate angular densities presented in Section~\ref{sec:gamspectral} are now fitted to the pseudo-sample of extreme high temperatures. The effects of the explanatory covariates $t$, $z$, and $d$ are tested in each of the three angular densities: the logistic model (Example~\ref{ex:logistic_surf}) with parameter $\alpha(t,z,d)$, the Dirichlet model (Example~\ref{ex:Dirichlet_surf}) with parameters $\alpha(t,z,d)$ and $\beta(t,z,d)$, and the H\"{u}sler--Reiss model (Example~\ref{ex:Husler_surf}) with parameter $\lambda(t,z,d)$. Within each family of covariate-adjusted angular densities, likelihood ratio tests (LRT) are performed to select the most adequate VGAM for the dependence parameters. Table \ref{VGAM_max_temp} shows the best models in each of the three families of angular densities.

\begin{table}[h!]
\caption{\footnotesize Selected models in each family of angular densities along with their AICs. The link functions $g$ are the logit function for the logistic model and the logarithm function for the Dirichlet and the H\"{u}sler--Reiss models. The functions $\hat{f}$ with subscripts $t$, $z$, and $d$ are fitted smooth functions of time, NAO, and day in season, respectively.}
  \centering
  \begin{tabular}{lcc}
  \hline   \hline
    Covariate-adjusted angular density & VGAM & AIC\\
    \hline
    Logistic &  $\hat{\alpha}(t,z,d)= g^{-1} \lbrace \hat{\alpha}_0+ \hat{f_t}(t)+\hat{f_z}(z)+\hat{f_d}(d)\rbrace $ & $-280.15$\\
    \multirow{2}{4em}{Dirichlet} & $\hat{\alpha}(z)= g^{-1} \lbrace \hat{\alpha}_0 + \hat{f_z}(z) \rbrace $ & \multirow{2}{4em}{ $-290.05$} \\ 
& $\hat{\beta}(t,d)= g^{-1} \lbrace \hat{\beta}_0 + \hat{f_t}(t) + \hat{f_d}(d) \rbrace $ & \\
    H\"{u}sler--Reiss & $\hat{\lambda}(t,z,d)= g^{-1} \lbrace \hat{\lambda}_0 + \hat{f_t}(t) + \hat{f_z}(z) + \hat{f_d}(d) \rbrace $ & $-275.64$\\
    \hline
  \end{tabular}
  \label{VGAM_max_temp}
\end{table}

All the considered covariates have a significant effect on the strength of dependence between extreme high temperatures in Montana and Zermatt. For the covariate-dependent Dirichlet model, the covariates affect the dependence parameters $\alpha$ and $\beta$ differently. However, these parameters lack interpretability, and \cite{Coles1994} mention the quantities $(\alpha+\beta)/2$ and $(\alpha-\beta)/2$ that can be interpreted as the strength and asymmetry of the extremal dependence, respectively. In this case, the best Dirichlet dependence model found in Table \ref{VGAM_max_temp} is such that both the intensity and the asymmetry of the dependence are affected by time, NAO, and day in season.

\begin{figure}
  \centering
  \subfloat{\includegraphics[width=\textwidth]{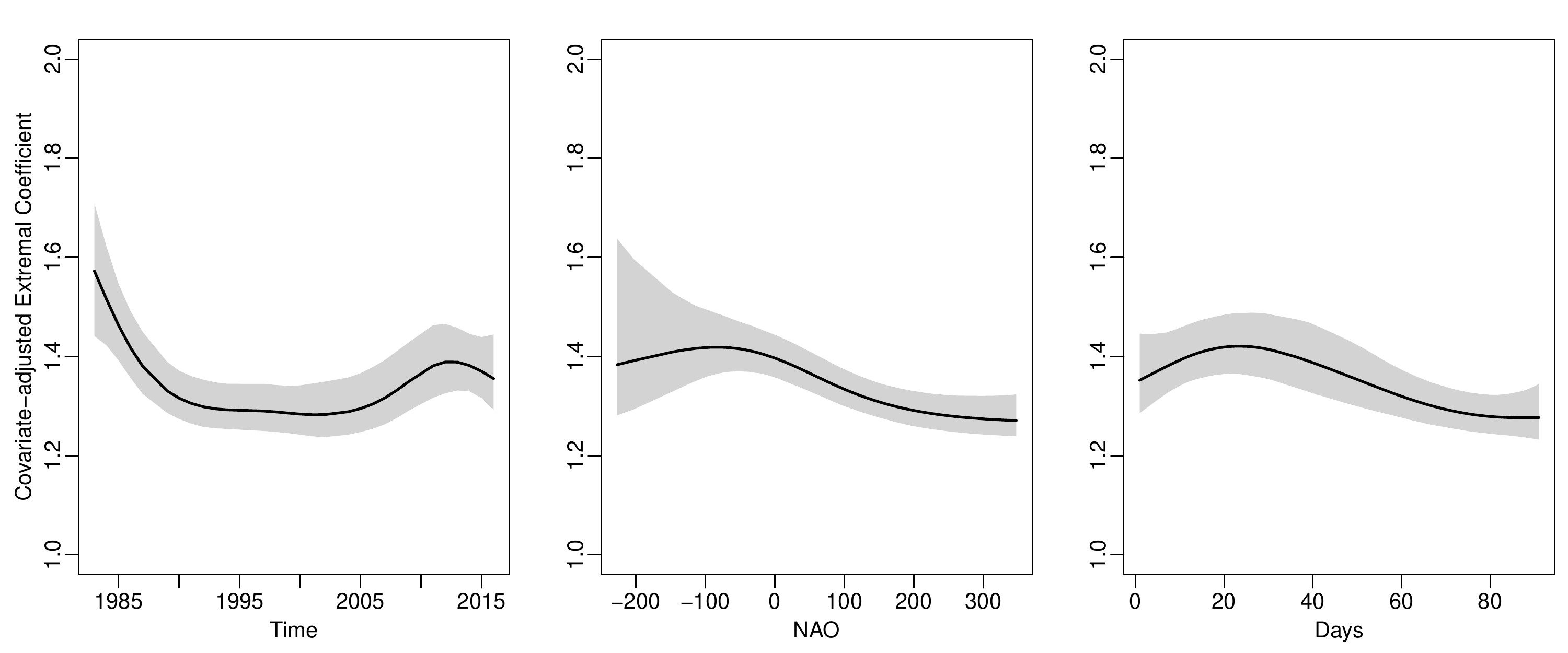}}
  \caption{\footnotesize Fitted smooth effects for the extremal coefficient under the Dirichlet model of Table \ref{VGAM_max_temp} along with their associated $95\%$ asymptotic confidence bands.\label{smooth_max_temp}}
\end{figure}

The best models in the studied angular density families are then compared by means of the AIC (see Section~\ref{subsec:selection_gamma}) displayed in Table \ref{VGAM_max_temp}. The Dirichlet model with $\alpha(z)$ and $\beta(t,d)$ parameters has the lowest AIC and is hence selected. This suggests the presence of asymmetry in the dependence of extreme high temperatures between Montana and Zermatt. 
Figure \ref{smooth_max_temp} shows the fitted smooth effects of the covariates on the extremal coefficient---constructed via the covariate-adjusted extremal coefficient as in \eqref{covext}---that lies between $1$ for perfect extremal dependence and $2$ for perfect extremal independence.

A decrease in the extremal coefficient, or equivalently an increase in the extremal dependence between high winter temperatures in Montana and Zermatt, is observed from $1988$ until $2006$. This change might be explained first by a warm phase of very pronounced and persistent warm anomalies during the winter season, which occured countrywide from $1988$ to $1999$ \citep{Jungo2001}, and then by an exceptionally warm $2006/2007$ winter that took place in Europe \cite{Luterbacher2007}. Regarding the NAO effect, as expected, we observe an increase in the extremal dependence during the positive phase of NAO that has a geographically global influence on the Alps and results in warmer and milder winters, as depicted by \cite{Beniston_1997}. In terms of the very negative NAO values (less than $-100$), there is an important uncertainty due to the corresponding small amount of joint extreme high temperatures (8\%). 

The right panel of Figure \ref{smooth_max_temp} suggests an increase in the extremal dependence around mid-December. This evidence also seems compatible with the countrywide findings by \cite{Beniston_1997}, who claims that
\begin{quote}\footnotesize
 ``The anomalously warm winters have resulted from the presence of very persistent high pressure episodes which have occurred essentially during periods from late Fall to early Spring.''
\end{quote}

\subsubsection*{Dependence of Extreme Low Winter Temperatures}
The effects of the covariates time, NAO, and day in season on the dependence between extreme cold winters in Montana and Zermatt are now tested by fitting the bivariate angular densities of Section~\ref{sec:gamspectral}. Within each of the logistic, Dirichlet, and H\"{u}sler--Reiss families, LRTs are performed, and the selected models are displayed in Table~\ref{VGAM_min_temp}.

\begin{table}[h!]
\caption{\footnotesize Selected models in each family of angular densities along with their AICs. The link functions $g$ are the logit function for the logistic model and the logarithm function for the Dirichlet and the H\"{u}sler--Reiss models. The functions $\hat{f}$ with subscripts $t$ and $d$ are fitted smooth functions of time and day in season, respectively.}
  \centering
  \begin{tabular}{lcc}
  \hline \hline
    Covariate-adjusted angular density & VGAM & AIC\\
    \hline
    Logistic &  $\hat{\alpha}(d)= g^{-1} \lbrace \hat{\alpha}_0+\hat{f_d}(d)\rbrace $ & $-402.76$\\
    \multirow{2}{4em}{Dirichlet} & $\hat{\alpha} \equiv g^{-1} ( \hat{\alpha}_0 )$ & \multirow{2}{4em}{ $-404.95$} \\ 
& $\hat{\beta}(t,d)= g^{-1} \lbrace \hat{\beta}_0 + \hat{f_t}(t) + \hat{f_d}(d) \rbrace $ & \\
    H\"{u}sler--Reiss & $\hat{\lambda}(t,d)= g^{-1} \lbrace \hat{\lambda}_0 + \hat{f_t}(t) + \hat{f_d}(d) \rbrace $ & $ -402.98$\\
    \hline
  \end{tabular}
  \label{VGAM_min_temp}
\end{table}

The explanatory covariates have different effects on the extremal dependence, depending on the family of angular densities. The AICs for the fitted models are quite close, and the asymmetric Dirichlet model has the lowest AIC and is hence the retained model. As opposed to the extremal dependence between warm winters in the two mountain sites, the NAO has a non-significant effect on the extremal dependence between cold winters. This might be explained by the fact that high values of the NAO index will affect the frequency of extreme low winter temperatures (less extremes) and hence the marginal behavior of the extremes at both sites, but not necessarily the dependence of the extremes between these sites \cite[sec.~7.3.2]{Beniston_2004}.

\begin{figure}
  \centering
  \subfloat{\includegraphics[width=\textwidth]{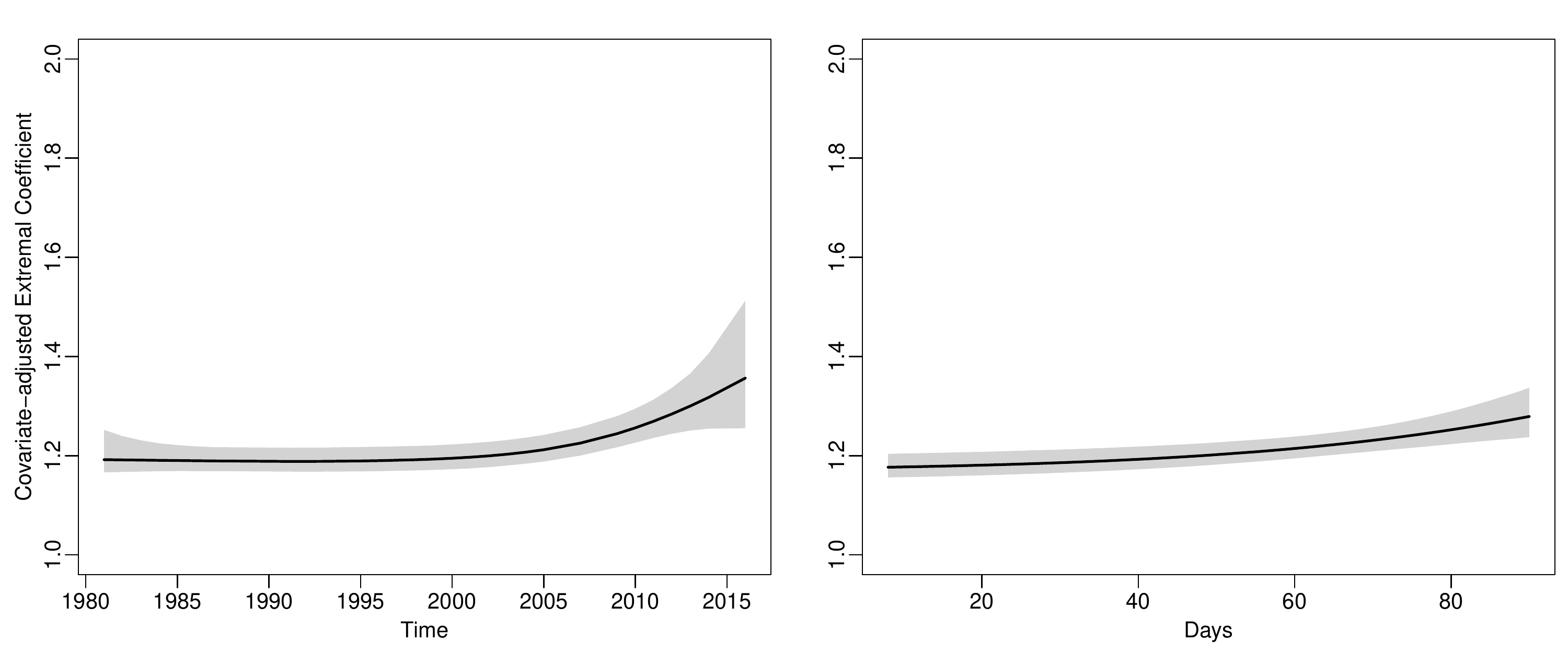}}
  \caption{\footnotesize Fitted smooth effects for the extremal coefficient under the Dirichlet model of Table \ref{VGAM_min_temp} along with their associated $95\%$ asymptotic confidence bands.\label{smooth_min_temp}}
\end{figure}

Figure \ref{smooth_min_temp} shows the fitted smooth effects of time and day in season. The extremal dependence between low winter temperatures in Montana and Zermatt is high, regardless of the values taken by the covariates $t$ and $d$. The range of values of the extremal coefficient observed in Figure~\ref{smooth_min_temp} is in line with the findings of \cite{Davison_Huser_Thibaut}, where the value of the extremal coefficient for the dependence between extreme low winter temperatures (in Switzerland) is around $1.3$ for pairs of resorts separated by up to $100$km. Overall, the extremal coefficient is lower in the extreme low winter temperatures than in the extreme high winter temperatures. This could be explained by the fact that minimum winter temperatures are usually observed overnight when the atmosphere is purer and not affected by local sunshine effects and hence is more favorable to the propagation over space of cold winter spells.

\begin{figure}
  \centering
  \subfloat{\includegraphics[width=0.5\textwidth]{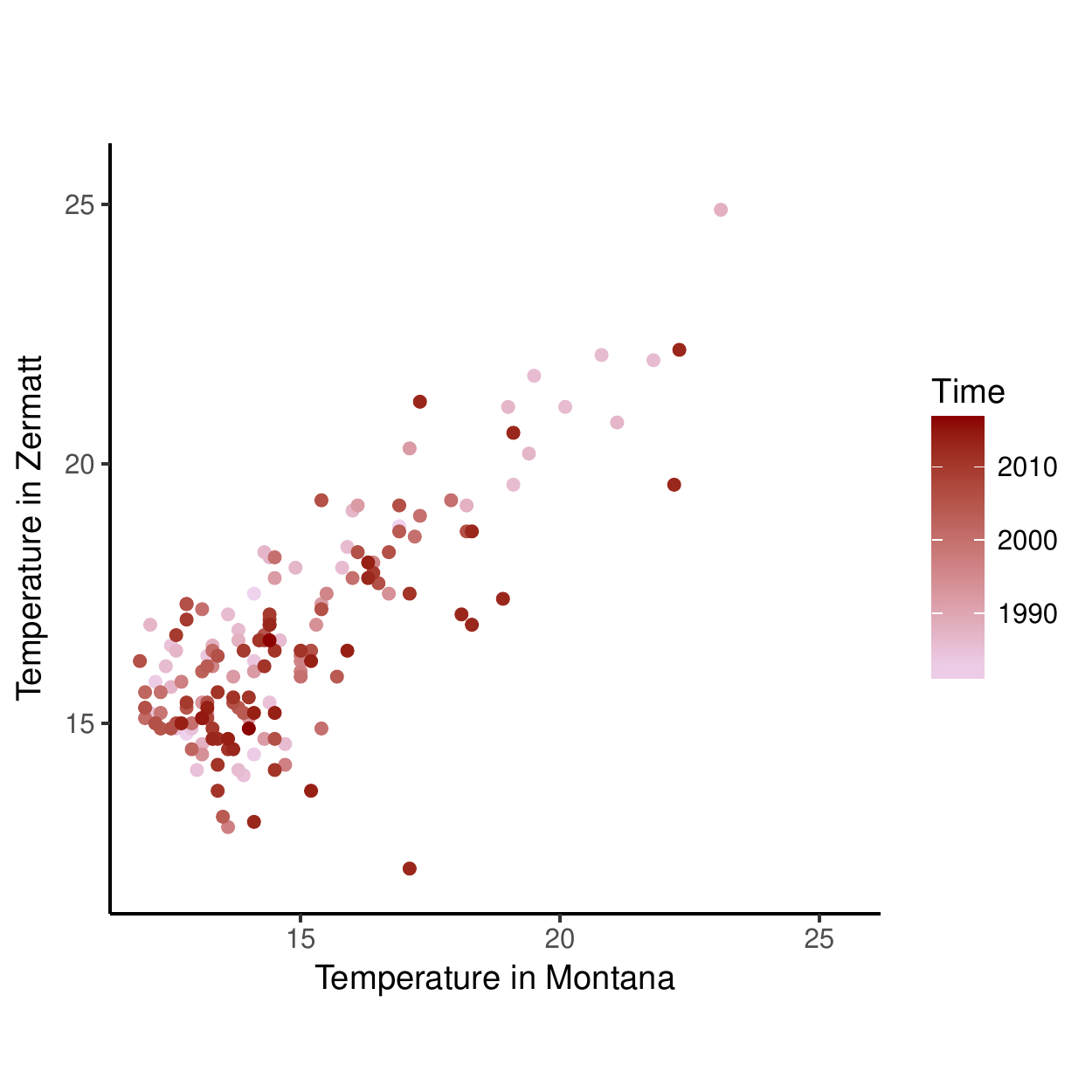}}
  \caption{\footnotesize Scatterplot of (minus) extreme low winter temperatures (in$\ ^{\circ}$C) in Montana and Zermatt.\label{min_temp_joint}}
\end{figure}

A decrease in the extremal dependence is observed from around $2007$ and results in values of the extremal coefficient that are comparable to those obtained under the warm winter spells scenario (see Figure \ref{smooth_max_temp}). This can be explained by a decrease in the intensity of the joint extreme low temperatures, that is, milder joint extreme low temperatures, occurring during the last years of the analysis, as can be observed in Figure~\ref{min_temp_joint}. The right panel of Figure~\ref{smooth_min_temp} highlights a decrease in the extremal dependence when approaching spring. This effect can be explained by the fact that mountains often produce their own local winds.\footnote{{\tt https://www.morznet.com/morzine/climate/local-climate-in-the-alps}} These warm dry winds are mostly noticeable in spring and are called Foehn in the Alps. Local effects obviously lead to a decrease of extremal dependence between the two resorts.

\section{\textsf{Final Remarks}}
\label{sec:discuss}
\noindent
In this paper, we have introduced a sturdy and general approach to model the influence of covariates on the extremal dependence structure. Keeping in mind that extreme values are scarce, our methodology borrows strength from a parametric assumption and benefits directly from the flexibility of VGAMs. Our non-linear approach for covariate-varying extremal dependences can be regarded as a model for conditional extreme value copulas---or equivalently as a model for nonstationary multivariate extremes. An important advantage over existing methods is that our model profits from the VGAM framework, allowing the incorporation of a large number of covariates of different types (continuous, factor, etc) as well as the possibility for the smooth functions to accommodate different shapes. The fitting procedure is an iterative ridge regression, the implementation of which is based on an ordinary N--R type algorithm that is available in many statistical software. An illustration is provided in the \R~code in the Supplementary Materials.


The method paves the way for novel applications, as it is naturally tailored for assessing how covariates affect dependence between extreme values---and thus it offers a natural approach for modeling conditional risk. Conceptually, the proposed approach is valid in high dimensions. Yet, as for the classical setting without covariates, the number of parameters would increase quickly with the dimension and additional complications would arise. Relying on composite likelihoods \citep{Padoan} instead of the full likelihood seems to represent a promising path for future extensions of the proposed methodology in a high-dimensional context. 


\bigskip
\noindent {\large\bf \textsf{Supplementary Materials}}

\noindent The online supplement to this article contains supplementary numerical experiments, \R~codes for implementing VGAM family functions for different angular density families, as well as the \R~codes used for the extreme temperature analysis.

\begin{description}
\item[Monte Carlo Evidence:] The file contains the results of the Monte Carlo study conducted in Section~\ref{montecarlo}. (.pdf file)
\item[Covariate Adjusted Angular Densities:] The file contains \R~codes for implementing the following angular density VGAM families: the bivariate logistic, the bivariate Dirichlet, the bivariate H\"{u}sler--Reiss, and the trivariate pairwise beta (see Section~\ref{sec:gamspectral}). Examples of the use of the implemented VGAM families are provided. (.zip file)
\item[Temperature Data Analysis:] The file contains the datasets obtained from the MeteoSwiss website as well as the \R~codes for the analysis of the extremal dependence between winter temperatures in Montana and Zermatt. (.zip file)
\end{description}

\if0\blind
{
\bigskip
\noindent {\large\bf \textsf{Acknowledgments}} \\ 

\vspace{-.5cm}\noindent We thank the Editor, Associate Editor, and two anonymous referees for several insightful recommendations that substantially improved the paper.
} \fi
\if1\blind
{
\bigskip
\noindent {\large\bf \textsf{Acknowledgments}} \\
\noindent We thank the Editor, Associate Editor, and two anonymous referees for several insightful recommendations that substantially improved the paper. We extend our thanks to the participants of Workshop 2017, EPFL, for discussions and comments, and to Paul Embrechts for his constant encouragement.

\bigskip
\noindent {\large\bf \textsf{Funding}} \\
The research was partially funded by FCT (Funda\c c\~ao para a Ci\^encia e a Tecnologia, Portugal) through the project UID/MAT/00006/2013. 
} \fi
\renewcommand\refname{\textsf{References}} 
\bibliographystyle{asa2.bst}  
\bibliography{mybib} 



\end{document}